\documentclass[prl,twocolumn,secnumarabic,amsmath,amssymb,
nofootinbib,superscriptaddress]{revtex4-1}

\usepackage{graphics}      
\usepackage{graphicx}      
\usepackage{longtable}     
\usepackage{url}           
\usepackage{bm}            
\usepackage{epsfig}
\usepackage{subfigure}
\usepackage[titletoc]{appendix}
\usepackage{wrapfig}
\usepackage{algorithmic,algorithm}
\usepackage{ifthen}
\usepackage{enumitem}
\usepackage{float}
\usepackage{mdframed}
\usepackage{paralist}

\usepackage{todonotes}
\usepackage{xcolor,colortbl}
\usepackage{soul}

\graphicspath{{./figs/}}

\usepackage[pagebackref=false,bookmarks=false]{hyperref} 

\definecolor{dblue}{rgb}{0.03,0.3,0.62}
\definecolor{dorange}{rgb}{1,0.55,0}

\setstcolor{red}

\hypersetup{
  bookmarksnumbered=true,
  bookmarksopen=false,
  hypertexnames=false,     
  breaklinks=true,          
  unicode=false,
  pdffitwindow=true,        
  pdfnewwindow=true,        
  colorlinks=true,         
  linkcolor=dblue,
  anchorcolor=red,
  citecolor=dblue,
  filecolor=magenta,
  urlcolor=dblue,
  pdfstartview = FitH,
  pdfkeywords = {},
  pdfcreator = {LaTeX with hyperref package}
}

\definecolor{sblue}{cmyk}{0.98,0.13,0,0.43} 
\definecolor{sblue}{cmyk}{0.98,0.13,0,0.43} 

\newcommand{\figref}[1]{Fig.~\ref{#1}}

\newcommand{\algref}[1]{Algorithm \ref{#1}}

\newcommand{\huback}{\widehat{\mathbf{u}}_{\infty,k}}

\newcommand{\GG}{\mathbf{G}}

\newcommand{\uback}{\mathbf{u}_{\infty}}

\newcommand{\uself}{\mathbf{u}_{\mathrm{self}}}

\newcommand{\calT}{\mathcal{T}}

\newcommand{\calB}{\mathcal{B}}

\newcommand{\calP}{\mathcal{P}}
\newcommand{\calM}{\mathcal{M}}
\newcommand{\calS}{\mathcal{S}}
\newcommand{\calR}{\mathcal{R}}
\newcommand{\XX}{\mathbf{X}}
\newcommand{\calX}{\mathcal{X}}

\newcommand{\tilxx}{\widetilde{\mathbf{x}}}
\newcommand{\ten}{\sigma}
\newcommand{\tenself}{\sigma_{\mathrm{self}}}

\newcommand{\UU}{{\mathbf{U}}}
\newcommand{\VV}{{\boldsymbol{V}}}
\newcommand{\WW}{{\boldsymbol{W}}}
\newcommand{\SSigma}{{\boldsymbol\Sigma}}

\newcommand{\xx}{{\mathbf{x}}}

\newcommand{\ee}{{\mathbf{e}}}

\newcommand{\yy}{{\mathbf{y}}}

\newcommand{\ff}{{\mathbf{f}}}

\newcommand{\CC}{{\mathbf{C}}}

\newcommand{\cc}{{\mathbf{c}}}
\newcommand{\uu}{{\mathbf{u}}}

\newcommand{\rr}{{\mathbf{r}}}
\newcommand{\nn}{{\mathbf{n}}}

\newcolumntype{C}{>{\centering\arraybackslash} m{2.5cm}}

\begin{document}
\title{Machine learning acceleration of simulations of Stokesian suspensions}

\author         {G\"{o}kberk Kabacao\u{g}lu}
\email          {gokberk@ices.utexas.edu}
\affiliation    {Department of Mechanical Engineering, The University of Texas at Austin, Austin, TX, 78712, United States}
\author         {George Biros}
\email          {gbiros@acm.org}
\affiliation    {Department of Mechanical Engineering, The University of Texas at Austin, Austin, TX, 78712, United States}
\affiliation    {The Oden Institute for Computational Engineering and Sciences, The University of Texas at Austin, Austin ,TX, 78712, United States}


\begin{abstract}
Particulate Stokesian flows describe the hydrodynamics of rigid or deformable particles in Stokes flows. Due to highly nonlinear fluid-structure interaction dynamics, moving interfaces, and multiple scales, numerical simulations of such flows are challenging and expensive. In this Letter, we propose a generic machine-learning-augmented reduced model for these flows. Our model replaces expensive parts of a numerical scheme with multilayer perceptrons. Given the physical parameters of the particle, our model generalizes to arbitrary geometries and boundary conditions without the need to retrain the regression function. It is 10$\times$ faster than a state-of-the-art numerical scheme having the same number of degrees of freedom and can reproduce several features of the flow quite accurately. We illustrate the performance of our model on integral equation formulation of vesicle suspensions in two dimensions.

\end{abstract}

\maketitle

\textit{Introduction.\textemdash}Particulate Stokesian flows consider the motion of a collection of rigid or deformable particles (e.g., drops, capsules, cells, slender bodies, filaments, active swimmers, possibly elastic or  filled by a fluid) that are suspended in a Newtonian fluid and the particle Reynolds number  is vanishingly small~\cite{lauga-powers09,shelley-zhang11,happel2012low,biesel2012}. Such flows find many applications in industrial processes,  microfluidics, study of complex fluids, bacterial and general active flows. Due to the multiple scales, the strongly nonlinear and nonlocal coupling of the interface deformation to the background flow, and the need for long time horizons, numerical simulations, no matter what the underlying numerical method is, can be extremely expensive. In this letter, we propose a \emph{machine-learning-augmented reduced model \textbf{(MLARM)}} that consists of three components:
\begin{inparaenum}[\bfseries{} (i)]
 \item We use multilayer perceptrons \textbf{\textit{(MLPs)}} to approximate several spatial nonlinear operators in our numerical scheme.  We choose these operators based on their stiffness properties, computational costs, and ability to generalize to unseen data. 
 \item We use  high-fidelity (i.e., highly refined in space and time) simulations to train the MLPs. We run these high-fidelity simulations in ``burst mode'', single-particle simulations for one time step. These simulations are done only once per particle type (e.g., for various mechanical properties of a particle) and target accuracy (i.e., time step size). 
 \item We combine the MLPs with low-fidelity simulations. \emph{Although we trained using single-particle, short-horizon, unconfined flows, our method enables us to conduct long-horizon simulations of suspensions with several particles in confined geometries.}
\end{inparaenum}
%

The basic idea is to create a regression function that accurately captures the dynamics of the flow using the high-fidelity simulations; and then use this function in the low-fidelity numerical scheme. There exist many schemes for function approximation in high dimensions; we  have opted for a multilayer perceptron~\cite{hornik-white-e89}. We model different aspects of a flow using separate MLPs. We train a MLP on single particle dynamics in a particle relaxation regime (no imposed flow); and we train another one for the evolution of boundary due to imposed velocity using the mode decomposition of its restriction on the boundary of the particle. Once we have trained the MLPs, the framework can be applied to \emph{any flow configuration}, that is, we can vary the confined geometry, the number of particles, and the imposed boundary conditions (confined or free-space) \emph{without the need to retrain the networks}. We demonstrate MLARM's capability of accurately capturing microscopic and macroscopic flow characteristics to the low-resolution simulations having the same number of degree of freedom for several benchmark problems.

A popular and effective methodology for the mathematical analysis and numerical simulation of Stokesian particulate flows is the boundary integral method (see~\cite{rallison-acrivos78,pozrikidis92,pozrikidis01a,wang-guan-e13,freund14} for various examples). Our overall method is based on this underlying formulation. To demonstrate our framework, we select a specific particle type, \emph{vesicle}~\cite{seifert97,kantsler-steinberg-e05,misbah12,li-karniadakis-e13}. (Although the details depend on a particle type, our model is broadly applicable to boundary integral formulations for Stokesian particulate flows). These flows are quite challenging because vesicles are  deformable particles that resist bending and stretching, and are filled with a fluid. Vesicles resemble red blood cells and their simulations have been used to help understanding microcirculation and to design microfluidic devices for medical diagnoses and drug delivery systems. Nevertheless, the high computational cost makes  simulations of vesicles at realistic concentrations extremely expensive (sometimes they can take several days on large clusters) and limits the ability to perform design optimization and parameter sweeps. Although a numerical scheme based on low resolution discretization in space and time along with a set of correction algorithms enables simulating dense vesicle flows for long time horizons~\cite{quaife-biros14,kabacaoglu-biros-e18} and solving a design optimization problem~\cite{kabacaoglu-biros18b}, such simulations still demand even faster algorithms~\cite{zhao-freund-e10,freund14,malhotra-biros-e17}. Let us note that although numerous works have used machine learning to tackle computational physics problems there has been little work on Stokesian flows. Most closely related to ours have proposed a reduced order model for unsteady quasi-one dimensional Euler flows~\cite{wang-ray-e19}. Reduced basis functions are extracted from high-fidelity solutions and deep neural networks are used to map the flow parameters to the coefficients of the reduced basis functions. \emph{But, to our knowledge there exist no works on reduced models for Stokesian particulate flows.}

\begin{figure}
\begin{minipage}{0.22\textwidth}
\setcounter{subfigure}{0}
\centering
\renewcommand*{\thesubfigure}{(a)} 
\vspace{1.2cm}\hspace{-0.5cm}\subfigure[A vesicle in a free-space flow]{\scalebox{0.20}{{\includegraphics{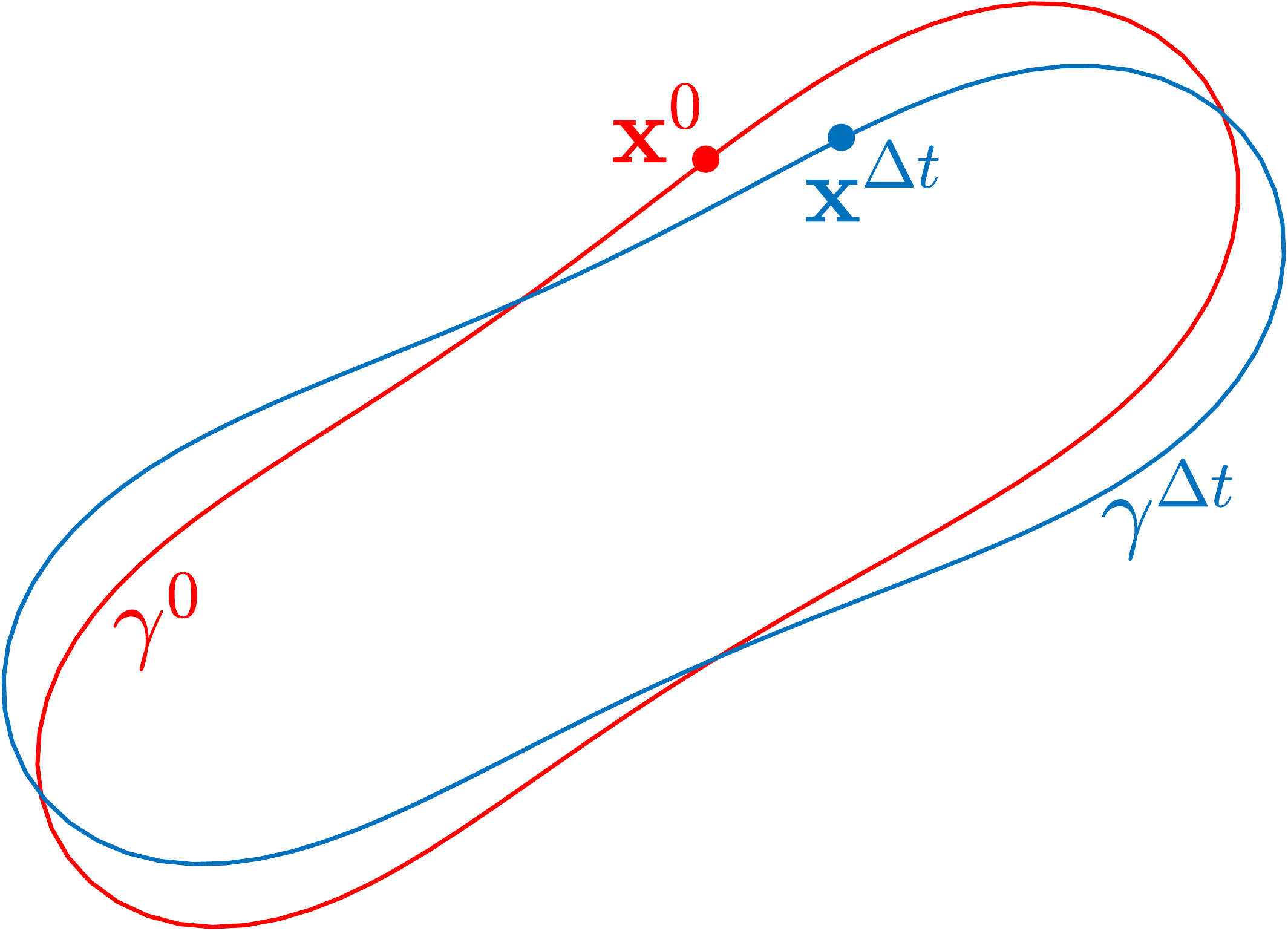}}}
\label{f:singleVes}} 
\end{minipage}
\begin{minipage}{0.22\textwidth}
\setcounter{subfigure}{0}
\centering
\renewcommand*{\thesubfigure}{(b)} 
\hspace{0.25cm}\subfigure[Vesicles in a confined flow]{\scalebox{0.24}{{\includegraphics{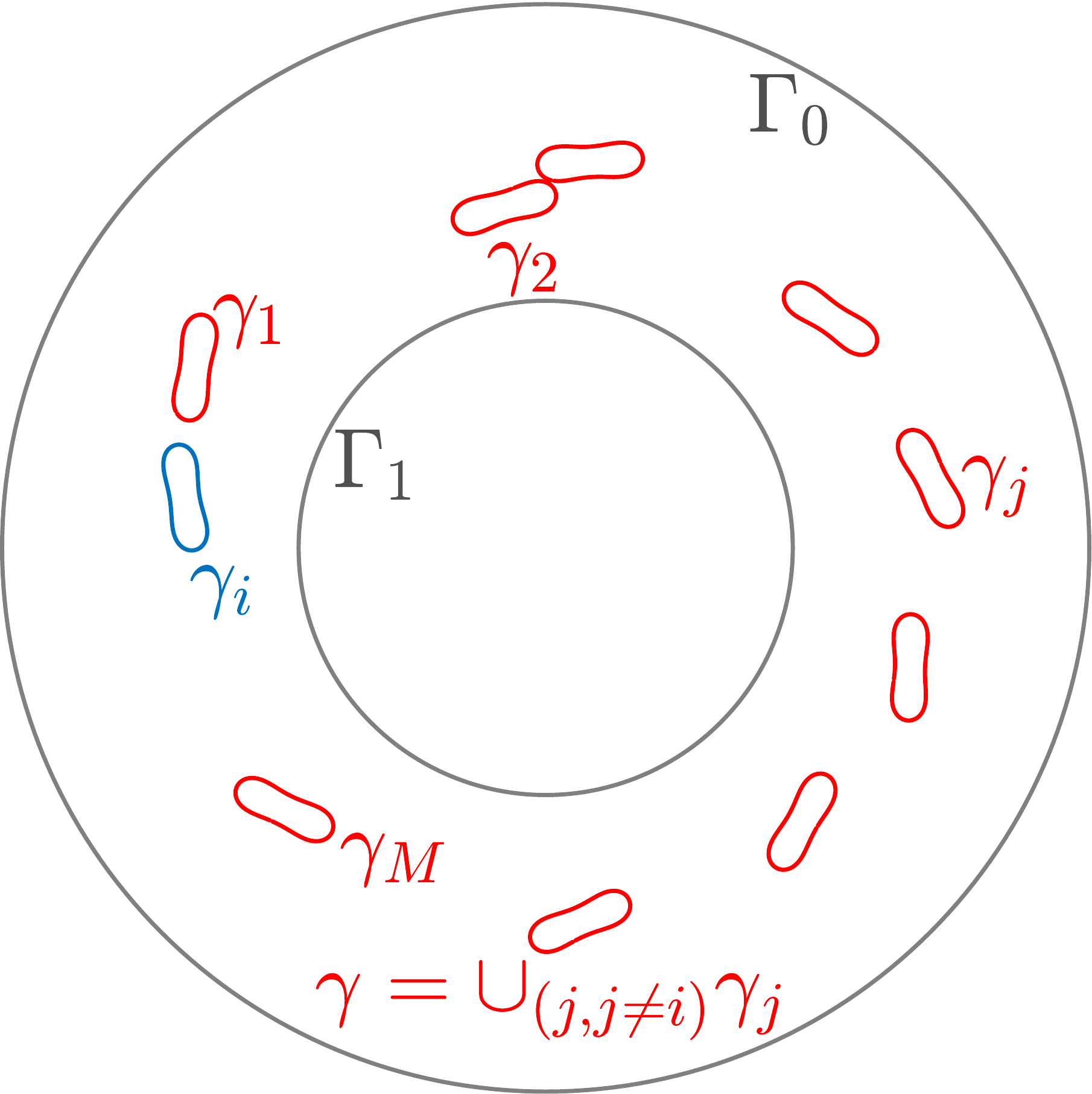}}}
\label{f:vesWwalls}} 
\end{minipage}
\caption{Problem setup and notation. We denote a vesicle membrane and a point on a membrane with $\gamma$ and $\xx$, respectively. (a) shows the evolution of $\gamma^{0}$ (red) into $\gamma^{\Delta t}$ (blue) after a time step. In (b), $\Gamma$ stands for fixed boundaries (i.e., solid walls) (gray), $\gamma_i$ and $\gamma$ represent the $i^{th}$ membrane (blue) and the other membranes (red).}
\label{f:probGeometry}
\end{figure}

\textit{MLARM for vesicle flows.\textemdash}We assume that fluids in the interior and the exterior of  the vesicles are Newtonian with the same viscosity. We consider two-dimensional flows and model them using a quasi-static Stokes approximation scheme~\cite{kraus-lipowsky-e96,danker-misbah-e09}.  Consider a single vesicle in a free-space flow (\figref{f:singleVes}). Let $\gamma$ denote a vesicle membrane. The points $\xx(\alpha,t)$ on $\gamma$ are given at $N$ uniformly distributed points $\{\alpha_k = 2\pi(k-1)/N\}_{k=1}^N$ in the parametric domain and time $t$. The velocity of the membrane points is given by
\begin{equation}\label{eq:aVesInFreeSpace} 
\frac{\partial \xx}{\partial t} = \uself[\gamma](\xx) + V[\gamma]\uback(\xx),
\end{equation}
where $\uself[\gamma](\xx)$ is the velocity induced by the vesicle itself and $V[\gamma]\uback(\xx)$ is the matrix-vector multiplication that evaluates the effect of the imposed flow $\uback$ on the membrane velocity. We refer to matrices with upper case letters and to vectors with lower case bold letters; $[\gamma]$ denotes dependence (nonlinear and nonlocal) on the interface $\gamma$, and it involves solution of boundary value problems (in our case with integral equations). 

Although in our prior work we have found linearly implicit time-stepping schemes to work well~\cite{kabacaoglu-biros-e18}, here we use a different method that allows modular MLP training. We use an operator-splitting method~\cite{macnamara-strang16} for~\eqref{eq:aVesInFreeSpace}, which brings several advantages when using MLPs (will appear later). The method divides \eqref{eq:aVesInFreeSpace} into two problems: advection and relaxation. Given the vesicle membrane $\gamma^0$, we first obtain the membrane $\gamma^*$ at the intermediate step by evaluating the membrane velocity due to the imposed flow, i.e., solving 
\begin{equation}\label{eq:advectFreeSpace} 
\frac{\partial \xx}{\partial t} = V[\gamma^0]\uback(\xx); \,\, \xx(0) = \xx^0 \in \gamma^0; \,\, t \in (0,\Delta t),
\end{equation}
where $\Delta t$ is the time step size. Then, the new membrane position $\xx^{\Delta t} \in \gamma^{\Delta t}$ is given by solving the relaxation problem 
\begin{equation}\label{eq:relaxFreeSpace} 
\frac{\partial \xx}{\partial t} = \uself[\gamma^*](\xx); \,\, \xx(0) = \xx^* \in \gamma^*; \,\, t \in (0,\Delta t).
\end{equation}
The computationally expensive steps in this scheme are constructing or applying $V[\gamma]$ and solving the relaxation problem~\eqref{eq:relaxFreeSpace}, which requires the expensive computation of $\uself[\gamma]$. Both operators depend nonlinearly on $\gamma$. In MLARM these steps are replaced with function approximations via MLPs as follows. First, for $V[\gamma]$, we need to remove the dependence on $\uback$ in order to enable generalization to unseen $\uback$ fields. We exploit the fact that $V[\gamma]\uback$ depends \emph{linearly} on $\uback$. We decompose $\uback$ using an $N_f$-term truncated Fourier series, $\uback(\xx(\alpha)) = \sum_{k=1}^{N_f}\phi_k(\alpha)\huback$, where $\phi_k$ are Fourier basis vectors and $\huback$ are the corresponding Fourier coefficients. Then, the term $V[\gamma]\uback(\xx)$ in~\eqref{eq:advectFreeSpace} becomes
\begin{equation*}\label{eq:ubackFourier} 
V[\gamma]\uback(\xx) = \sum_{k=1}^{N_f} V[\gamma]\phi_k(\alpha)\huback = \sum_{k=1}^{N_f}\Psi_k[\gamma](\alpha)\huback.
\end{equation*}
We approximate $\{\Psi_k[\gamma]\}_{k=1}^{N_f}$ using $N_f$ MLPs, which we term $V$-MLPs. Then, given an unseen $\gamma$ and $\uback$, we first compute the $\Psi_k$ operators using the $V$-MLPs and then apply the inverse Fourier transform. As a bonus, $\Psi_k[\gamma]$ turns out to be linear on certain  parameters such as the membrane's bending rigidity and the time step size, thus, there is no need to retrain the MLPs for these parameters. For the relaxation problem~\eqref{eq:relaxFreeSpace} we train another MLP that approximates the nonlinear function $\calR(\gamma^0) = \gamma^{\Delta t}$. We call this one $R$-MLP. The function $\calR$ is nonlinear in the problem parameters, therefore, retraining the MLP for different values of bending rigidity and time step size is needed. Choosing a time step size depends on viscous and bending forces on a vesicle. We train several $R$-MLPs for different values of time step size and bending rigidity. Given flow parameters we determine the appropriate time step size and choose the corresponding $R$-MLP. One can also build a parametric reduced model using the trained MLPs~\cite{amsallem-farhat08}.

For multiple vesicles in confined flows (\figref{f:vesWwalls}), only the advection problem~\eqref{eq:advectFreeSpace} changes. For the $i^{th}$ vesicle it becomes
\begin{equation}\label{eq:advectMultiCon} 
\frac{\partial \xx}{\partial t} = V[\gamma_i^0]\uback[\gamma,\Gamma](\xx); \,\, \xx(0) = \xx^0 \in \gamma_i^0,
\end{equation}
where $\uback$ is the velocity on $\gamma_i$ due to its interactions with the other membranes $\gamma$ and the fixed boundaries $\Gamma$. Computing $\uback$ requires solving additional integral equations: one for the density on the fixed boundaries and one for the inextensibility constraint. Boundaries do not evolve with time so the related matrices are precomputed and their application can be accelerated with fast multipole methods (FMM)~\cite{greengard-rokhlin87}. The inextensibility constraint requires solving an equation for Lagrange multiplier (tension). While this equation can be eliminated for a single vesicle, that is not the case for multiple vesicles and confined geometries. The equation resembles~\eqref{eq:aVesInFreeSpace} and we solve it using $N_f + 1$ MLPs. Once the tension for all vesicles and the density on the boundaries are obtained, we use FMM to compute $\uback[\gamma,\Gamma]$. Then we compute the term $V[\gamma_i]\uback[\gamma,\Gamma]$ with the approximated $\Psi_k[\gamma]\huback$ using $N_f$ MLPs. Lastly, the relaxation problem~\eqref{eq:relaxFreeSpace} is solved.  Evaluating layer potentials at points close to a boundary (fixed or vesicle) requires special quadratures~\cite{pozrikidis99,kropinski99,helsing-ojala08,zhao-freund-e10,quaife-biros14}. These methods can be quite expensive to be implemented in MLARM. So, we correct near interactions between boundaries by  employing a kinematic collision handling that is also used in simulations of emulsions~\cite{zinchenko-davis08} and red blood cells~\cite{zhao-freund-e10} in 3D. For the minimum arclength spacing $h$ as a threshold, whenever the distance of a membrane point $\xx$ to another membrane is less than $h$, the point is moved in the $(\xx_p-\xx)$ direction, where $\xx_p$ is the projection of $\xx$ on the other membrane until $\|\xx-\xx_p\| = h/2$. To have smooth perturbations in vesicle membranes, we also move the neighboring points in the same direction but for smaller amounts. See the Supplemental Material for all the details on the formulation and MLARM. 

Overall, the MLPs approximate the computationally expensive terms that depend only on a vesicle shape $\gamma$. Our judicious application of MLPs allows us to train them only for various vesicle shapes, which eases data generation and use the trained MLPs for any flow configuration. 

\begin{figure}
\includegraphics[width=0.45\textwidth]{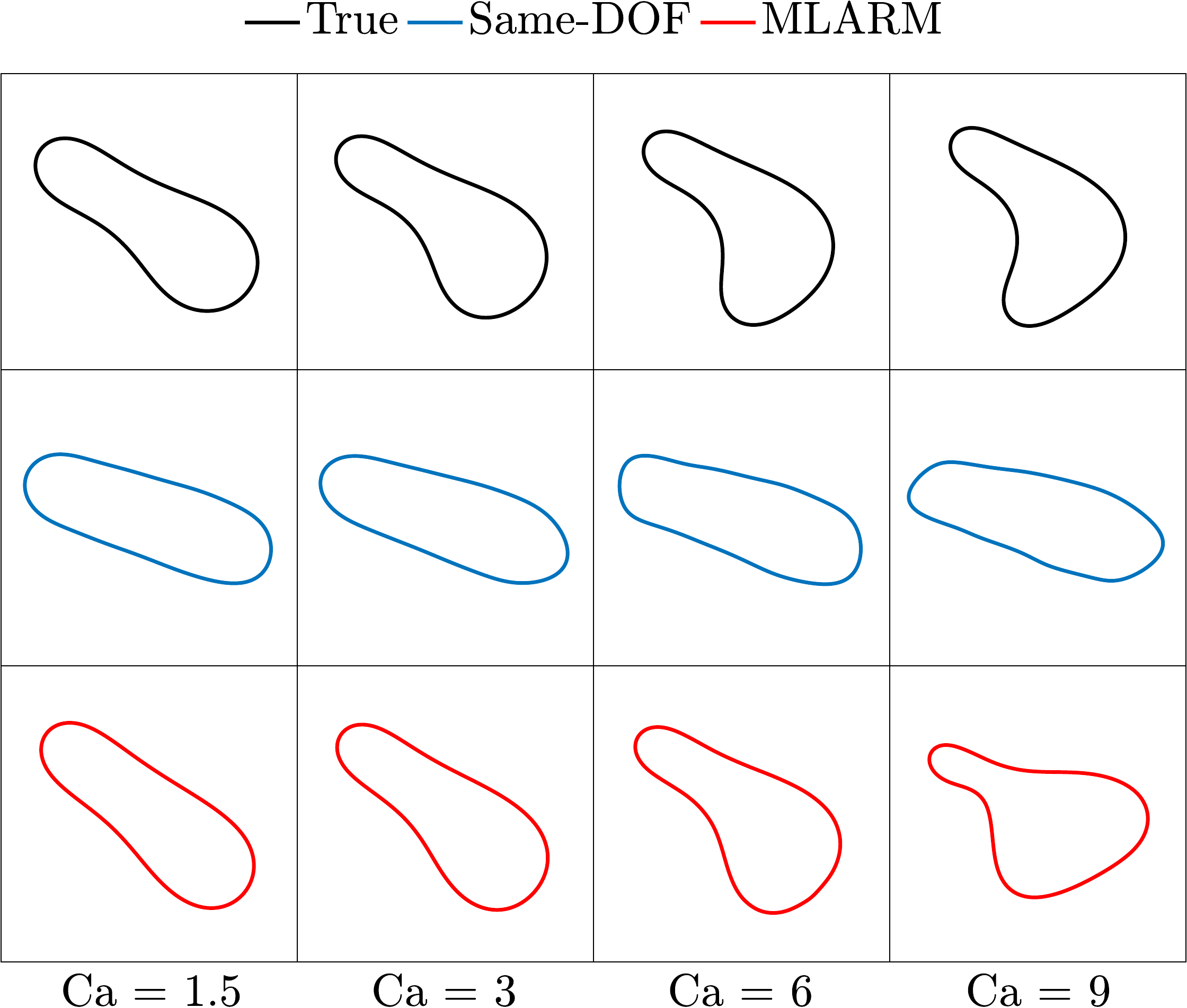}
\caption{Equilibrium vesicle shapes in a parabolic flow. The MLARM simulations are much more accurate than the same-DOF simulations in capturing the equilibrium shapes. See the Supplemental Material for a movie for the case of  Ca = 6.}
\label{f:parabolic}
\end{figure}

\textit{Training MLPs.\textemdash}For the MLPs we use a lower dimensional representation of a vesicle shape based on \textit{principal component analysis (PCA)}. Given a training set of vesicle shapes, we find a reduced modal basis for the shapes using PCA and use it to represent the input and output to the MLPs. 
Approximately 100,000 distinct shapes make up the library. The outputs are high-fidelity solutions of the related integral equations. For $R$-MLP we solve the relaxation problem~\eqref{eq:relaxFreeSpace} for only one time step to obtain $\gamma^{\Delta t}$. The terms approximated by the MLPs linearly depend on translation, rotation and scaling of a vesicle. So, we standardize vesicle shapes so that they have the same center, angular orientation and arclength.
The MLPs consist of five to six fully-connected layers with approximately 75,000 to 85,000 parameters. See the Supplemental Material for details on the training data and the MLP's architectures.

\begin{figure}
\begin{minipage}{0.22\textwidth}
\setcounter{subfigure}{0}
\centering
\renewcommand*{\thesubfigure}{(a)} 
\hspace{0cm}\subfigure[True simulation]{\scalebox{0.25}{{\includegraphics{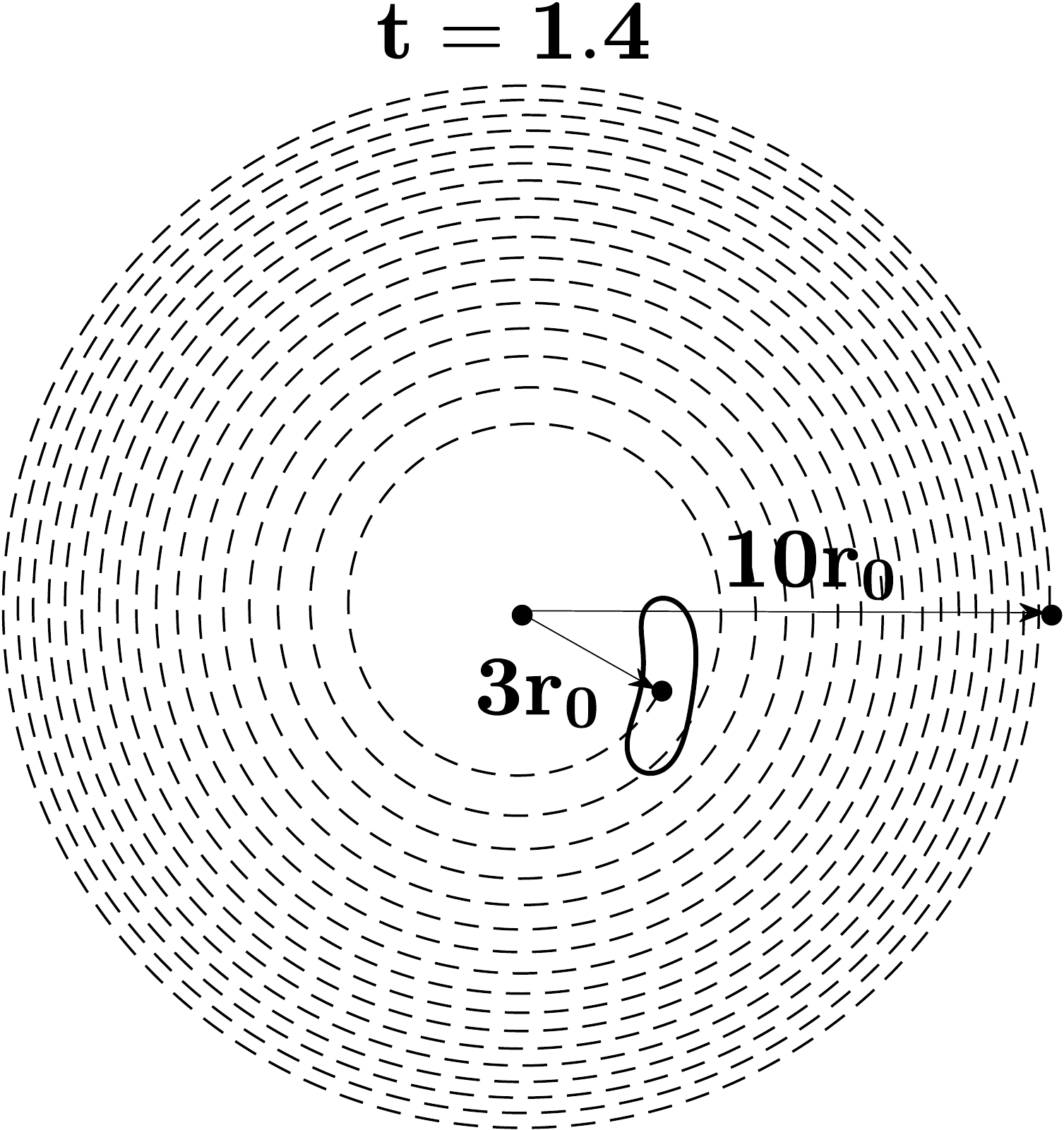}}}
\label{f:rotTrue}} 
\end{minipage}
\begin{minipage}{0.22\textwidth}
\setcounter{subfigure}{0}
\centering
\renewcommand*{\thesubfigure}{(b)} 
\hspace{0.5cm}\subfigure[MLARM simulation]{\scalebox{0.25}{{\includegraphics{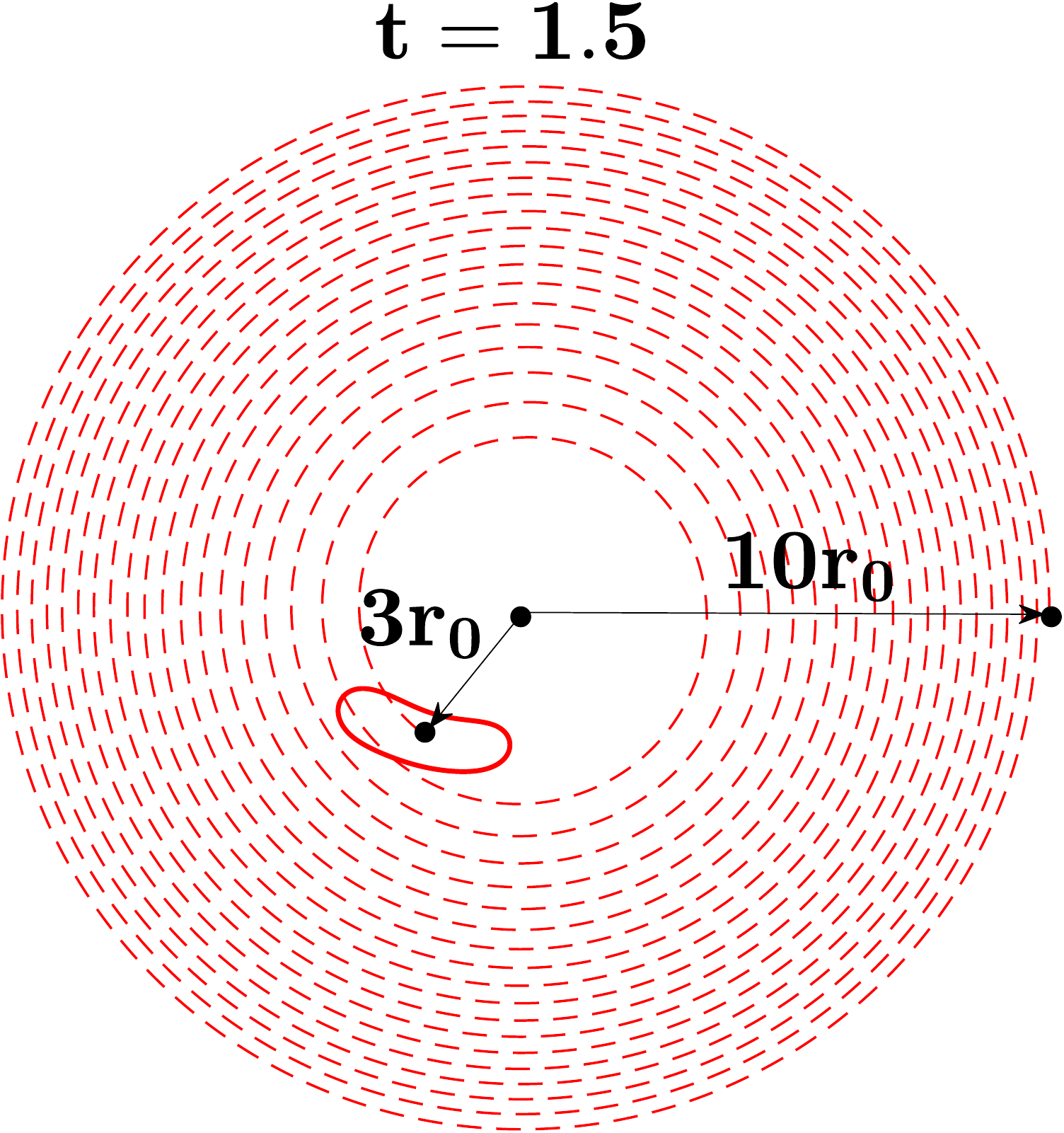}}}
\label{f:rotMLP}} 
\end{minipage}
\caption{Vesicle trajectories in a flow with curved flow lines. A vesicle initialized at $10r_0$ migrates towards the center in time $t$. The final radial position is $3r_0$. The MLARM simulation accurately captures the migration with 6\% error in migration velocity. See the Supplemental Material for the movie.}
\label{f:rotation}
\end{figure}
\begin{figure*}
\includegraphics[width=0.95\textwidth]{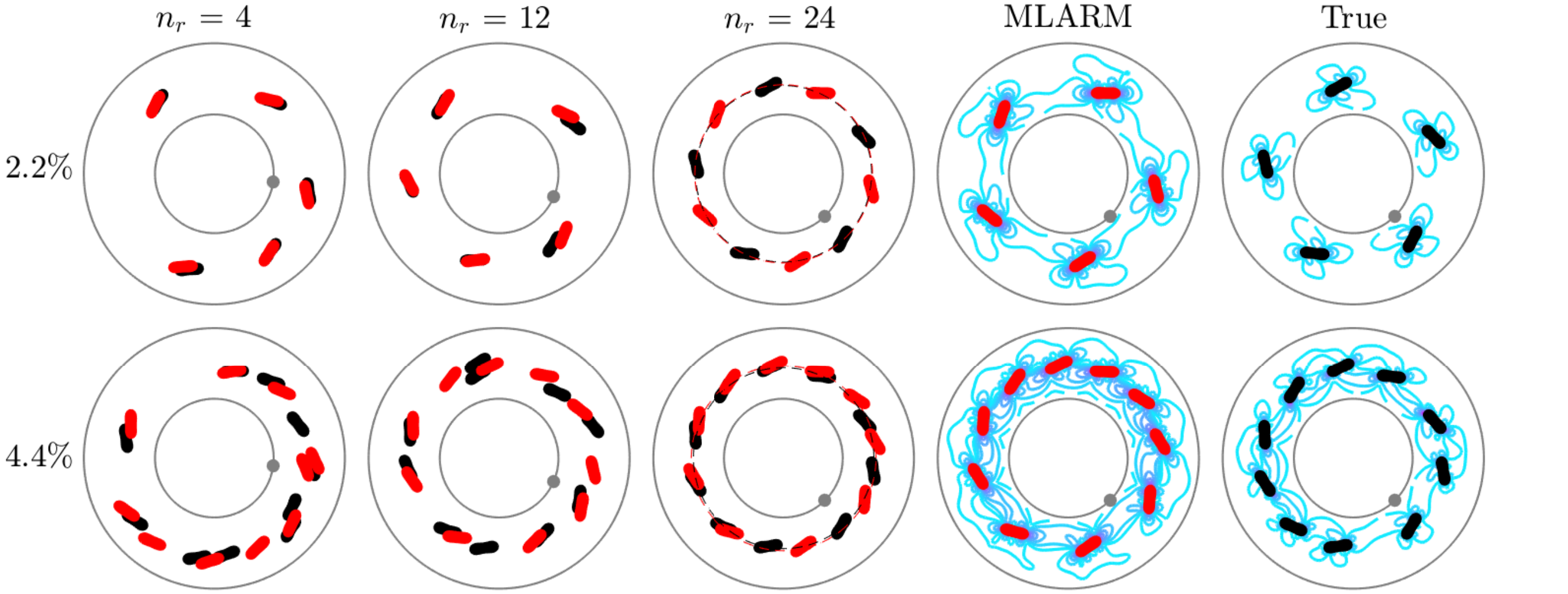} \caption{Equilibrium organization
in dilute suspensions in a Taylor-Couette flow (area fractions of 2.2\% at the top and 4.4\% at the bottom). The MLARM and true solutions are superimposed and shown in red and black, respectively. At the equilibrium of such flows, vesicles exhibit a spatial order by moving in a rim with a uniform angular interdistance. We present
snapshots after $n_r$ number of rotations of the inner circle. The dashed lines in the third column show the equilibrium rims. The last two columns demonstrate the magnitude of the perturbation in the velocity field
induced by vesicles after $n_r = 24$ rotations. See the Supplemental Material for the movies.} \label{f:couette} 
\end{figure*}
\textit{Numerical experiments.\textemdash}We demonstrate MLARM's capabilities  on several benchmark vesicle flows. We measure length in units of the vesicle radius $r_0$ (defined as the radius of a circle having the same enclosed area). The flows are characterized by capillary number $\mathrm{Ca} = \mu r_0^3 \dot{\gamma}/\kappa_b$ where $\mu$ is the dynamic fluid viscosity and $\dot{\gamma}$ is the imposed shear rate that is varied to adjust Ca. \textit{True} solutions are obtained using high-fidelity simulations~\cite{quaife-biros14}. We compare the MLARM simulations and \textit{the same-degree-of-freedom (DOF) simulations} performed with the numerical scheme that has the same number of DOF as MLARM in terms of accuracy and computation time. We also compare with \textit{the same-cost simulations} that have the same computation time with the MLARM simulations in terms of accuracy.

\begin{figure}
\includegraphics[width=0.4\textwidth]{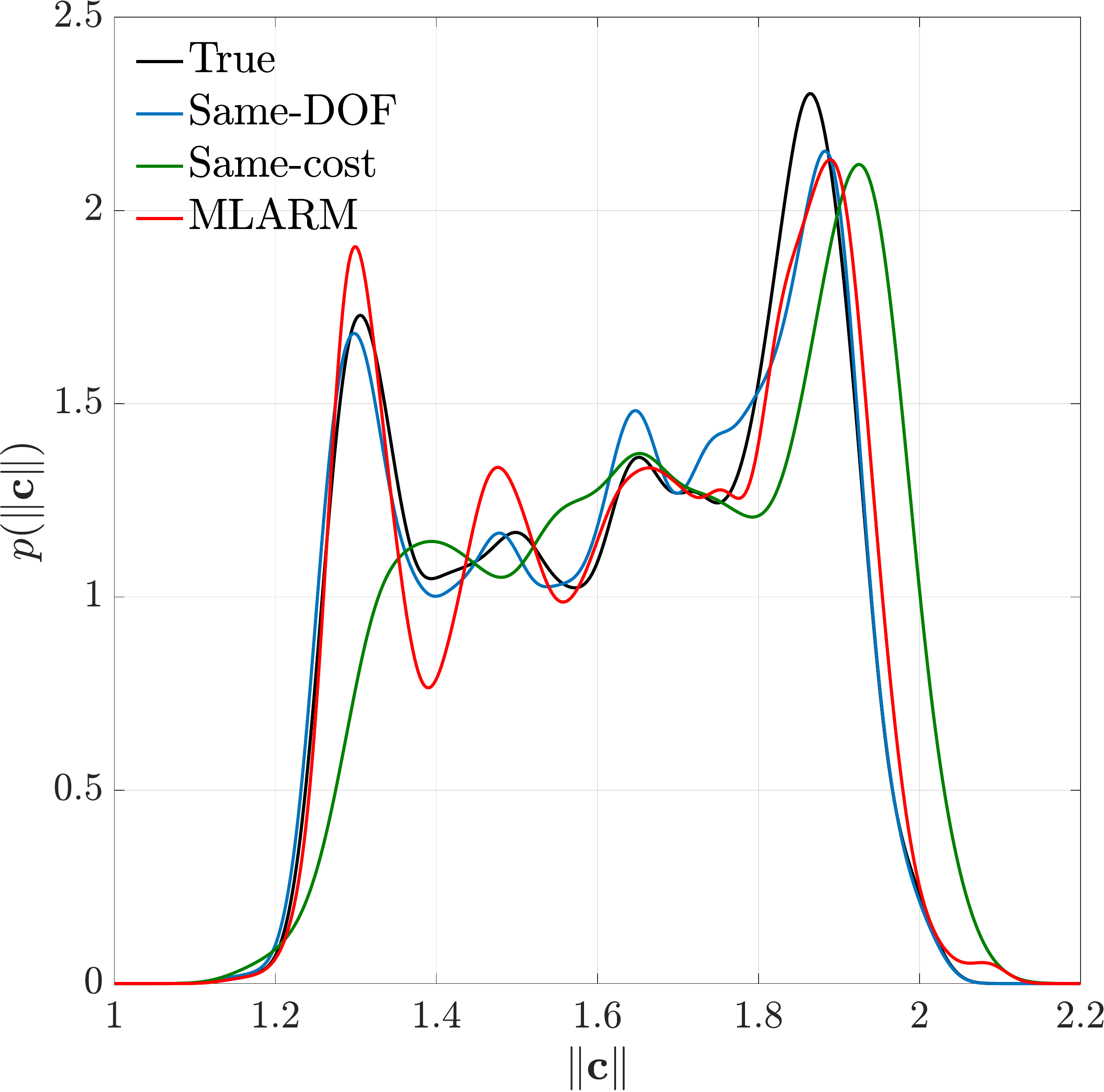} \caption{Statistics of vesicles in a dense Taylor-Couette flow at an area fraction 20\%. We plot the probability distribution of the distance of the vesicle center to the origin $\| \mathbf{c}\|$ throughout the simulation. Vesicles form regions near circles that are free of vesicles. The MLARM simulation accurately captures these regions. See the Supplemental Material for the movies.} \label{f:VF20stats} 
\end{figure}
First, we consider a vesicle initialized at $y = r_0/2$ in a parabolic flow ($\uback = r_0\dot{\gamma}\left(1-(y/W)^2\right)\ee_x$) with $W = 10r_0$. In this flow a vesicle migrates towards the low shear rate region and reaches an equilibrium shape depending on its reduced area and Ca. We reproduce the results for a vesicle of reduced area 0.65 in~\cite{kaoui-misbah09}. The equilibrium shape is asymmetric (slipper-like) for all Ca values. Figure~\ref{f:parabolic} shows that the MLARM simulations capture the true equilibrium shapes for all $\mathrm{Ca} < 9$ more accurately than the same-DOF simulations do. For $\mathrm{Ca} = 9$, the scheme requires finer temporal resolution. For this case, the MLARM simulations are slightly faster than the same-DOF ones. See the Supplemental Material for the vesicles' equilibrium lateral positions and the comparison with the same-cost simulations. The MLARM simulations are more accurate in capturing the equilibrium lateral positions than the same-cost ones while they have similar accuracy in the equilibrium shapes.

Second, we study cross-streamline migration of a vesicle suspended in a flow with curved flow lines. The setup is the same as in~\cite{ghigliotti-misbah-e10}. The imposed shear rate is $\dot{\gamma} = 20/r^2$ and the velocity field is $(v_{\theta},v_r) = (-10/r,0)$ where $r$ is the distance between the vesicle's center and the origin. The value of Ca depends on $r$ (varying from $\mathrm{Ca} = 0.2$ at $r = 10r_0$ to $\mathrm{Ca}\approx 2.2$ at $r = 3r_0$). In this flow a vesicle with properties we consider (no viscosity contrast and reduced area of 0.65) migrates towards regions of high shear rates (\figref{f:rotTrue}). We measure migration velocity in the radial direction from $r = 10r_0$ to $r = 3r_0$. Figure~\ref{f:rotMLP} shows that the MLARM simulation captures the migration with the error $|v_{\mathrm{mig}}^{\mathrm{MLA}}-v_{\mathrm{mig}}^{\mathrm{True}}|/v_{\mathrm{mig}}^{\mathrm{True}} = 0.06$. We do not present the same-DOF and the same-cost simulations as they are completely wrong. 

Third, we consider dilute suspensions in Taylor-Couette flow. Here the inner circle rotates in the counterclockwise direction while the outer one is stationary. The distance between the inner and the outer circles is $10r_0$ and Ca = 1.5. In such flows, the force applied by the inner circle balances vesicles' inward migration. The vesicles eventually organize themselves in a rim with the same interdistance for area fractions of vesicles between approximately 1\% and 4\%~\cite{ghigliotti-misbah-e10}. We perform simulations with area fractions 2.2\% and 4.4\% until the inner circle completes 24 rotations. Figure~\ref{f:couette} shows that closely and randomly initialized vesicles reach to approximately the same radial positions and are separated uniformly in the azimuthal direction at the equilibrium. The MLARM simulations can capture the vesicles' spatial order although the individual trajectories are inaccurate.  The MLARM simulation becomes less accurate than the same-DOF simulation as area fraction increases since MLARM ignores the near vesicle interactions (see the Supplemental Material for the evolution of the mean radial positions).  Note that resolving individual trajectories is not of interest in suspensions because of their sensitivity to (generally unknown) initial conditions.

Finally, we consider dense suspensions in a Taylor-Couette flow. The setup is the same as in the previous example. Here vesicles migrate away from the circles and form regions near circles that are free of vesicles (called cell-free layers). We investigate how accurately MLARM can capture these layers. We plot the probability distribution of the distances of the vesicles' centers to the origin throughout the simulations in~\figref{f:VF20stats} for an area fraction 20\%.  The MLARM and the same-DOF simulations accurately capture the layers and the same-cost simulation is less accurate. See the Supplemental Material for the results of area fractions 30\% and 35\% and also the movies. There is no resolution that provides stable and the same computational cost as the MLARM simulation in the 30\% and 35\% area fraction cases. Here the MLARM simulations are 10$\times$ faster than the same-DOF simulations and 25$\times$ faster than the ground truth simulations. This example clearly shows the effectiveness of the MLARM simulations compared to using just low-resolution simulations.

\textit{Conclusion.\textemdash}We propose a reduced model that combines multilayer perceptrons trained with high-fidelity simulations for one time step and a low-fidelity numerical scheme. In various examples of vesicle flows, we compare the MLARM simulations with those given by the numerical scheme with the same degree-of-freedom and also that with the same computational cost. MLARM is faster and also provides simulations that have either similar or better accuracy. One of the limitations of MLARM is that we have not optimized the MLPs' architectures and training parameters. Such optimization can provide more accurate approximations.  Although we presented our analysis in 2D flows (so that we can ensure highly accurate ground truth simulations), the model generalizes as is to 3D vesicle flows and other particles like deformable capsules, drops, filaments and rigid bodies.

\bibliography{main}

\begin{widetext}
\section{Supplemental Material}

\subsection{Formulation}

In this section, we present the mathematical modeling and the boundary integral equation formulation of vesicle flows in two dimensions. 

\subsubsection{Governing equations}

We assume that there are no external forces and the same Newtonian fluid with viscosity $\mu$ occupies the interior and the exterior of vesicles. $K$ and $M$ are the numbers of fixed boundaries (solid walls) and vesicles, respectively. We denote the membrane (boundary) of the $i^{th}$ vesicle with $\gamma_i$. Let $\Gamma_1,\cdots,\Gamma_{K-1}$ be walls surrounded by another wall $\Gamma_0$ (Fig. 1(b)). Since we consider flows in the limit of vanishing Reynolds number, the Stokes equations govern the fluid flow:
\begin{equation}\label{eq:stokesEqns}
-\mu\Delta \uu(\xx) + \nabla p(\xx) = 0, \quad \text{and} \quad \nabla \cdot \uu(\xx) = 0, \quad \xx \in \Omega \setminus (\bigcup_i \gamma_i),
\end{equation}
where $\Omega$ is $K$-ly connected domain of interest, $\uu$ is the fluid velocity field and $p$ is the pressure. We impose the no-slip boundary condition on membranes and the velocity Dirichlet boundary condition on the walls as 
\begin{equation}\label{eq:noSlip}
\uu(\xx,t) = \frac{\partial \xx}{\partial t}(t), \,\, \xx \in \{\gamma_i\}_{i=1}^M \quad \mathrm{and} \quad \uu(\xx,t) = \UU(\xx,t), \,\, \xx \in \{\Gamma_k\}_{k=0}^{K-1}.
 \end{equation}
Vesicle membranes are locally inextensible, i.e., each vesicle has a constant arclength. So, \eqref{eq:stokesEqns} is subject to the following constraint
\begin{equation}\label{eq:inextensible}
\xx_s \cdot \uu_s = 0, \,\, \xx \in \{\gamma_i\}_{i=1}^M.
\end{equation}
The subscript $s$ stands for differentiation with respect to the arclength. Finally, the momentum balance on a membrane requires the jump in surface traction to be equal to the membrane force due to bending and tension, i.e.,
\begin{equation}\label{eq:tracJump}
 [\![ \mathbf{T}\nn ]\!] = -\kappa_b \xx_{ssss} + (\sigma \xx_s)_s, \,\, \xx \in \{\gamma_i\}_{i=1}^M,
\end{equation}
where $\mathbf{T} = -p\mathbf{I} + \mu (\nabla \uu + \nabla \uu^T)$ is the Cauchy stress tensor, $\nn$ is the outward normal vector on $\gamma_i$, $ [\![ \cdot ]\!]$ is the jump across the membrane. The first term on the right hand side is the membrane force due to the bending rigidity $\kappa_b$ and the second one is the force due to the tension $\sigma$.

\subsubsection{Boundary integral equation formulation}

Let us first present the integral equation formulation of \eqref{eq:stokesEqns}-\eqref{eq:tracJump} for a single vesicle in a free-space flow, then, extend it to multiple vesicles in confined flows. The membrane force $\ff = -\kappa_b\xx_{ssss}+(\sigma\xx_s)_s$ induces velocity. The single layer integral $\calS[\ff](\xx)$ evaluates the velocity at $\xx$ as 
\begin{equation}\label{eq:SLP}
\calS[\ff](\xx) = \int_{\gamma}\GG(\xx-\yy)\ff\,d\gamma(\yy),
\end{equation}
where the free-space 2D Green's function for the Stokes equations is 
\begin{equation*}\label{eq:greensFunc}
\GG_{ij}(\xx-\yy) = \frac{1}{4\pi\mu}\left(-\delta_{ij}\ln|\xx-\yy|+\frac{(\xx-\yy)_i(\xx-\yy)_j}{|\xx-\yy|^2}\right).
\end{equation*}
For convenience, we introduce bending operator $\calB$, tension operator $\calT$, surface divergence operator $\calP$ and stretching operator $\calM$ defined for $\xx \in \gamma$ as follows
\begin{subequations}\label{eq:operators}
\begin{alignat}{1}
& \calB(\xx)\xx = -\kappa_b\calS[\xx_{ssss}]\xx, \label{eq:bendingOp}\\
& \calT(\xx)\sigma = \calS[\sigma\xx_s]\xx,  \label{eq:tensionOp} \\
& \calP(\xx)\yy = \xx_s\cdot\yy_s, \label{eq:surfDivOp} \\
& \calM(\xx) = \calT(\xx)\left(\calP(\xx)\calT(\xx)\right)^{-1}\calP(\xx).\label{eq:stretch}
\end{alignat}%
\end{subequations}
The velocity of the membrane points $\xx \in \gamma$ in a free-space flow with the velocity field $\uback$ is given as
\begin{equation}\label{eq:aVesIE}
\frac{\partial \xx}{\partial t} = (1-\calM(\xx))\calB(\xx)\xx + (1-\calM(\xx))\uback(\xx).
\end{equation}
So, the terms $\uself[\gamma](\xx)$ and $V[\gamma]$ in the letter (Eq. 1) correspond to $(1-\calM(\xx))\calB(\xx)\xx$ and $(1-\calM(\xx))$, respectively. Here the tension $\sigma$ is eliminated as the stretching operator modifies the membrane velocity to enforce the inextensibility constraint (see~\cite{shravan-biros-e09} for its derivation). 

For multiple vesicles in confined flows, the velocity of the points $\xx$ on the $i^{th}$ membrane $\gamma_i$ is given by the same equation as~\eqref{eq:aVesIE} with a different composition of the term $\uback$, i.e.,
\begin{equation}\label{eq:multiVesIE}
\frac{\partial \xx}{\partial t} = (1-\calM(\xx))\calB(\xx)\xx + (1-\calM(\xx))\uback[\gamma,\Gamma](\xx),
\end{equation}
where $\gamma = \cup_{j=1, j\neq i}\gamma_j$ and $\Gamma = \cup_{k=0}\Gamma_k$. The term $\uback$ is velocity induced by other vesicles in the flow and the walls on the $i^{th}$ membrane instead of the velocity field of the background flow as in~\eqref{eq:aVesIE}. The velocity $\uback[\gamma,\Gamma](\xx)$ is given as
\begin{equation}\label{eq:ubackMulti}
\uback[\gamma,\Gamma](\xx) = \sum_{\substack{j = 1 \\ j \neq i}}^M \int_{\gamma_j} \GG(\xx-\yy)\ff(\yy)\,d\gamma_j(\yy) + \cal{W}[\eta](\xx),
\end{equation}
where $\mathcal{W}[\eta]$ is the velocity induced by the walls due to the density $\eta$ on them. Computing~\eqref{eq:ubackMulti} requires solving integral equations for the tension $\sigma$ and the density $\eta$. The inextensibility constraint~\eqref{eq:inextensible} delivers the following equation for the tension
\begin{equation}\label{eq:tensionEqn}
\calP(\xx)\calT(\xx)\sigma(\xx) = -\calP(\xx)\left(\calB(\xx)\xx + \uback[\gamma,\Gamma](\xx)\right), \quad \xx \in \gamma_i.
\end{equation}
The Dirichlet velocity boundary condition on the walls provides the following equation for the density $\eta$ on $\Gamma$ 
\begin{equation} \label{eq:IEsolidWalls}
\UU(\xx) = -\frac{1}{2} \eta(\xx) + \sum_{j=1}^M\int_{\gamma_j} \GG(\xx-\yy)\ff(\yy)\,d\gamma_j(\yy) + \mathcal{W}[\eta](\xx), \quad  \xx \in \Gamma.
\end{equation}
See~\cite{rahimian-biros-e10} for the definition of the completed double layer integral $\mathcal{W}[\eta]$. We refer the reader to~\cite{rahimian-biros-e10} for the semi-implicit time stepping scheme and~\cite{quaife-biros14} for the implicit time stepping scheme to solve~\eqref{eq:aVesIE},~\eqref{eq:tensionEqn} and~\eqref{eq:IEsolidWalls}. We obtain the true solutions in the present study using the implicit time stepping scheme while our machine-learning-augmented reduced model is built based on the semi-implicit one.

\begin{figure}
\includegraphics[width=0.5\textwidth]{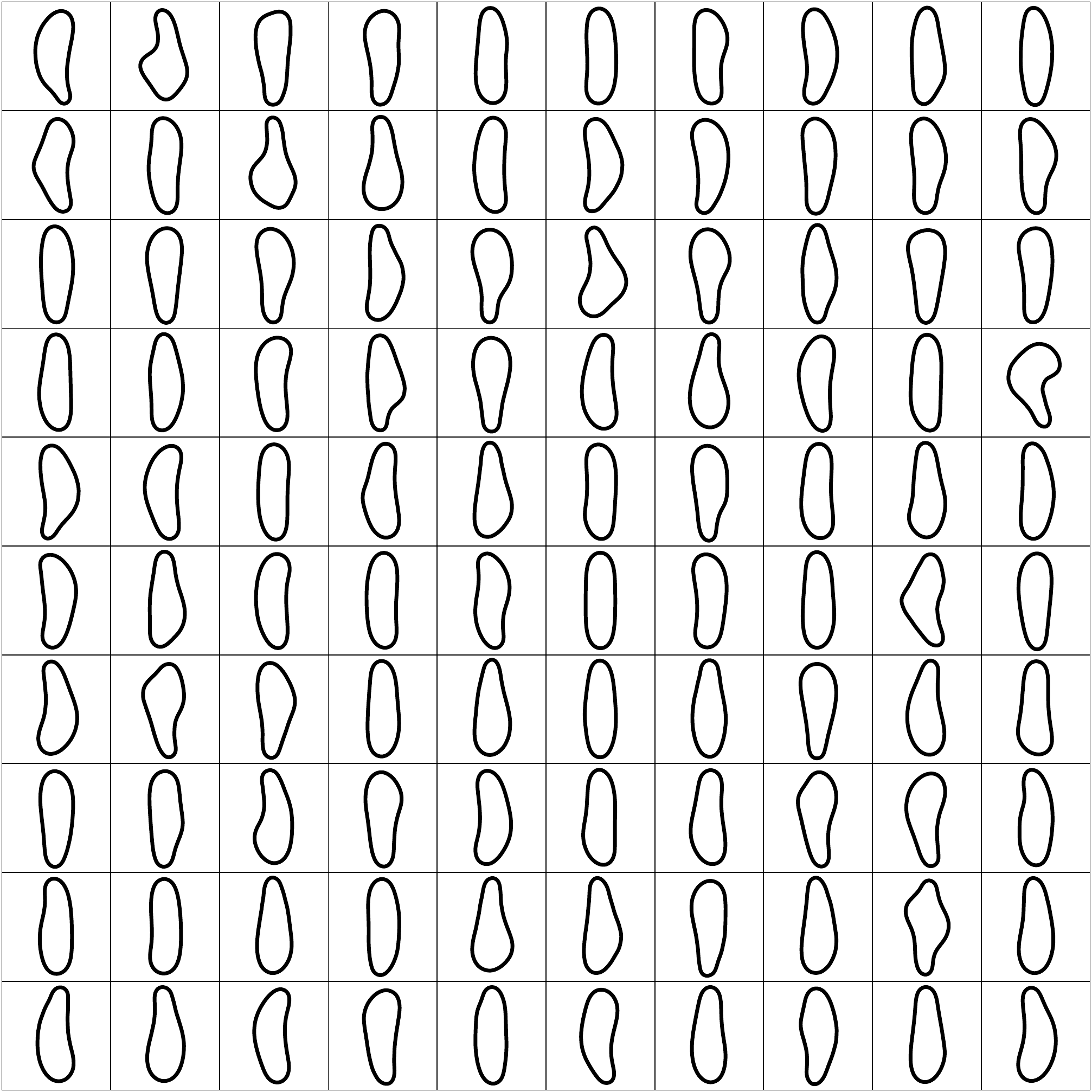}
\caption{Randomly sampled vesicles from the library. They are standardized such that their centers are at the origin, inclination angles are $\pi/2$ and arclengths are one.}
\label{f:vesLibrary}
\end{figure}
\subsection{Machine-learning-augmented reduced model}

In this section, we present the details of the machine-learning-augmented reduced model (MLARM). Specifically, we first explain the use of MLPs to obtain tension by solving~\eqref{eq:tensionEqn} for multiple vesicles in confined flows. Then, we elaborate on generating a data set, dimensionality reduction using principal component analysis (PCA), multilayer perceptrons' (MLP) architectures and parameters for training the MLPs. Finally, we present the pseudo-algorithm of MLARM.

\subsubsection{Using MLPs to obtain tension for flows of multiple vesicles and solid walls}

The way we enforce the inextensibility constraint requires solving~\eqref{eq:tensionEqn} to obtain tension for each vesicle at every time step. The solution for the $i^{th}$ membrane can be expressed as
\begin{equation}\label{eq:tensionForm}
\sigma(\xx) = \tenself[\gamma_i](\xx) + T[\gamma_i]\uback[\gamma,\Gamma], \quad \xx \in \gamma_i,
\end{equation}
where $\tenself[\gamma_i](\xx) = -(\calP(\xx)\calT(\xx))^{-1}\calP(\xx)\calB(\xx)\xx$ and $T[\gamma_i] = -(\calP(\xx)\calT(\xx))^{-1}$. So, the first term in~\eqref{eq:tensionForm} is the tension due to a vesicle itself and the second one is the contribution due to the other vesicles and the walls. The form of~\eqref{eq:tensionForm} resembles (1) for the velocity of the membrane points. That is why in MLARM we approximate $\sigma$ with a strategy similar to the one we use for (1). That is, we use a MLP to approximate the mapping between $\gamma_i$ and $\tenself[\gamma_i](\xx)$, and use $N_f$ MLPs for the mappings between $\gamma_i$ and $\{T[\gamma_i]\phi_k(s)\}_k^{N_f}$ where $\phi_k(s)$ is the Fourier basis vectors. We call the former $\ten$-MLP and the latter $T$-MLPs. The input to the MLPs is the PCA coefficients of a vesicle shape $\gamma_i$. Since the same PCA basis cannot be used to represent $\tenself[\gamma_i](\xx)$, the output of $\ten$-MLP is the Fourier coefficients corresponding to the first 32 modes which accurately reconstruct the term.

\subsubsection{Generating a data set}

Vesicles show wide variety of shapes in long time horizon simulations of their dense suspensions. In order to have various physical shapes we perform such a simulation of vesicles having the reduced area 0.65 in a confined Taylor-Couette flow (see~\cite{kabacaoglu-biros-e18} for the snapshots of the simulation). We want a library that consists of distinct vesicle shapes by some metric. In order to compare the shapes obtained from the simulation, we, first, standardize the shapes such that their centers, inclination angles\footnote{The inclination angle is the angle between the flow direction and the principal axis corresponding to the smallest principal moment of the inertia. The moment of inertia tensor is $J = \frac{1}{4}\int_{\gamma} (\rr \cdot \nn) (|\rr^2|I-\rr \otimes \rr)\,d\gamma(\xx)$ where $\rr = \xx-\cc$ and $\cc$ is the center of the vesicle.} and arclengths are the same. Then, we measure the dissimilarity between the standardized shapes based on the Hausdorff metric. Our library consists of $M = 100,181$ distinct vesicle shapes. See~\figref{f:vesLibrary} for 100 randomly sampled shapes from the library. 

Once we have the library, we generate sets of inputs and outputs for the MLPs using a high resolution discretization, i.e., $N = 96$ points per vesicle using~\algref{a:relaxAlgorithm}. As the approximated terms linearly depend on translation, rotation and scaling of a vesicle, we standardize the input vesicle shapes $\xx$ so that they have the same center and inclination angle. We also use a standard ordering of the discretization points on vesicle membranes. The first point is the one on the positive $x$-axis, then the other points are equally distributed along the arclength in the counterclockwise direction. The output of $R$-MLP is the solution of~\eqref{eq:aVesIE} with $\uback = 0$, the bending rigidity $\kappa_b = 1$ and the time step size $\Delta t = 10^{-4}$ only for one time step. This time step size is the largest one with which stable and accurate simulations in a stationary fluid can be performed using the semi-implicit time stepping scheme. For the other MLPs the outputs are the high-fidelity solutions of the related integral equations for vesicles in the library. We have found that $N_f = 24$ Fourier modes accurately represent the velocity $\uback$. Therefore, we use $N_f = 24$ MLPs to approximate the action of each $V$ and $T$ on the Fourier basis vectors ($V$-MLPs and $T$-MLPs). Each coefficient $k$ has two components, imaginary and real. Additionally, since we represent the vesicles with $N = 96$ and each point has two degrees of freedom, $V\phi_k$ and $T\phi_k$ are of sizes 192 and 96. We reduce this size by subsampling $V\phi_k$ to $N = 24$ points and $T\phi_k$ to $N = 48$. The outputs to the $k^{th}$ MLPs for these terms are vectors of size 96 that contain real and imaginary components of the terms. By choosing the subsampling rates as such we can use the same MLPs for both terms.

\begin{algorithm}[H]
\begin{algorithmic} 
\REQUIRE {Library of vesicle shapes, $\calX$}
\FOR[For every vesicle in $\calX$]{$\xx \in \calX$}
\STATE {$\xx = \mathtt{equallyDistributeInArcLength}(\xx)$}
\COMMENT {Equally distribute points along arclength}
\STATE {$[\tau,\theta,\mathtt{index}] = \mathtt{findStandard}(\xx)$}
\COMMENT {Find translation, rotation, ordering of points to standardize $\xx$ using Algorithm~\ref{a:findStandard}}
\STATE {$\xx_0 = \mathtt{standardize}(\xx, \tau,\theta,\mathtt{index})$}
\COMMENT {Standardize $\xx$ based on $(\tau,\theta,\mathtt{index})$ using Algorithm~\ref{a:standardize}}
\STATE {$[\calM,\calB,\calT,\calP] = \mathtt{solveIEs}(\xx_0,\kappa_b,\Delta t)$}
\COMMENT {Solve integral equations to build operators in~\eqref{eq:operators}}
\STATE {$\xx^{\Delta t} = \mathtt{implicitSolveRelax}(\xx_0,\kappa_b,\Delta t)$}
\COMMENT {Solve relaxation problem (\eqref{eq:aVesIE} with $\uback = 0$) implicitly}
\STATE {$ V = (1-\calM)$}
\COMMENT {Build velocity operator acting on $\uback$}
\STATE {$ \tenself = -(\calP\calT)^{-1}\calP\calB\xx_0 $}
\COMMENT {Find tension due to vesicle itself \eqref{eq:tensionForm}}
\STATE {$ T = -(\calP\calT)^{-1}$}
\COMMENT {Build tension operator acting on $\uback$ \eqref{eq:tensionForm}}
\STATE {$I(:,i) = \mathtt{map2reducedSpace}(\xx_0)$}
\COMMENT {Input to MLPs is PCA coefficients of $\xx_0$}
\STATE {$O_R(:,i) = \mathtt{map2reducedSpace}(\xx^{\Delta t})$} 
\COMMENT {Output to $R$-MLP is PCA coefficients of $\xx^{\Delta t}$}
\STATE {$O_V(:,:,i) = \{V\phi_k\}_{k=1}^{N_f}$}
\COMMENT {Output to $V$-MLPs is action of $V$ on Fourier basis vectors $\phi_k$}
\STATE {$O_{\ten}(:,i) = \mathtt{map2reducedSpace}(\tenself)$} 
\COMMENT {Output to $\ten$-MLP is Fourier coefficients of $\tenself$}
\STATE {$O_T(:,:,i) = \{T\phi_k\}_{k=1}^{N_f}$}
\COMMENT {Output to $T$-MLPs is action of $T$ on Fourier basis vectors $\phi_k$}
\ENDFOR
\end{algorithmic}
\caption{$[I,O_R,O_V,O_{\ten},O_T] = \mathtt{generateDataSet}(\calX)$, generates inputs and outputs for the MLPs} \label{a:relaxAlgorithm}
\end{algorithm}
\begin{algorithm}[H]
\begin{algorithmic} 
\REQUIRE {Coordinates of membrane points, $\xx = [x_1 \,\, x_2 \,\, \cdots \,\, x_N \,\, y_1 \,\, y_2 \,\, \cdots \,\, y_N]^T$}
\STATE {$\tau = -[\mathtt{mean}(\xx(1:N)) \,\, \mathtt{mean}(\xx(N+1:2N))]$}
\COMMENT {Find amount of translation to center $\xx$ at (0,0)}
\STATE {$\theta = \pi/2-\mathtt{getInclinationAngle}(\xx)$}
\COMMENT {Find amount of rotation to set inclination angle to $\pi/2$}
\STATE {$\mathtt{index} = \mathtt{findOrder}(\xx)$}
\COMMENT {Find ordering of points}
\end{algorithmic}
\caption{$[\tau,\theta,\mathtt{index}] = \mathtt{findStandard}$, finds translation, rotation, ordering of points to standardize a vesicle} \label{a:findStandard}
\end{algorithm}
\begin{algorithm}[H]
\begin{algorithmic} 
\REQUIRE {$\xx, \tau,\theta,\mathtt{index}$}
\STATE {$\xx = \mathtt{rotate}(\xx + \tau,\theta)$}
\COMMENT {First translate $\xx$ by $\tau$, then rotate it by $\theta$}
\STATE {$\xx_0 = \xx (\mathtt{index})$}
\COMMENT {Finally order indices of points}
\end{algorithmic}
\caption{$\xx_0 = \mathtt{standardize}$, translates and rotates a vesicle and order its discretization points} \label{a:standardize}
\end{algorithm}

\subsection{Principal component analysis}

We perform dimensionality reduction using principal component analysis. Here we describe how to form a reduced basis and represent a vesicle shape using this basis. A column vector $\xx = [x_1 \,\, x_2 \,\, \cdots \,\, x_N \,\, y_1 \,\, y_2 \,\, \cdots \,\, y_N]^T$ of size $2N$ contains the $x$ and $y$ coordinates of the membrane points and represents a vesicle shape $\gamma$. Let $\XX = [\xx_1 \,\, \xx_2 \,\, \cdots \,\, \xx_M]$ be a matrix of size $2N$-by-$M$ that stores $M$ vesicle shapes in the library. The singular value decomposition of $\XX$ is
\begin{equation*}
\XX = \WW \SSigma \VV^T
\end{equation*}
where $\WW$ is a $2N$-by-$2N$ matrix, $\SSigma$ is a $2N$-by-$M$ rectangular diagonal matrix of the singular values of $\XX$ and $\VV$ is an $M$-by-$M$ matrix. Let $\CC = \SSigma \VV^T$ be a $2N$-by-$M$ score matrix whose columns contain the representations of $M$ samples in the orthogonally transformed space. We obtain a truncated $\CC_{N_p}$ matrix of size $N_p$-by-$M$ by considering only the first $N_p$ largest singular values and the corresponding singular vectors,
\begin{equation*}
\CC_{N_p} = \WW^T_{N_p} \XX
\end{equation*}
where $\WW_{N_p} = [w_1 \,\, w_2 \,\, \cdots \,\, w_{N_p}]$ is an $2N$-by-$N_p$ matrix whose columns are the reduced basis vectors. Then the lower dimensional representation of a vesicle $\xx$ is $c = \WW^T_{N_p}\xx$. We set $N_p = 16$ as the first 16 PCA modes capture 99\% of the total energy, i.e., $(\sum_{i=1}^{16} \sigma_i^2)/ (\Sigma_{i=1}^{2N} \sigma_i^2) > 99\%$. 

\subsubsection{MLP architectures}

A detailed introduction of multilayer perceptrons is beyond the scope. We only introduce definitions that are helpful to our presentation of the MLPs' architectures (see~\cite{goodfellow-courville-e16} for more information). A MLP has a number of parameters that are found by minimizing a loss function. Let $c$ and $c^{\Delta t}$ be the lower dimensional representations of vesicles $\xx$ and $\xx^{\Delta t}$. Then, we find the parameters of $R$-MLP by minimizing the mean squared error 
\begin{equation*}\label{eq:lossBending}
\mathcal{J} = \sum_{j=1}^M \| \xx^{\Delta t}_j - \tilxx^{\Delta t}_j \|_2,
\end{equation*}
where $\tilxx^{\Delta t}_j = W_{N_p}\widetilde{c}^{\Delta t}_j$ correspond to the MLP's approximation to the $j^{th}$ vesicle in the data set of size $M$. $\| \cdot \|_2$ is the Euclidean norm. One can also measure the mean squared error based on the PCA coefficients $c$. We have also used a MLP trained to minimize such an error and obtained similar results. The parameters of the $k^{th}$ $V$-MLP ($T$-MLP as well) minimize the mean squared error 
\begin{equation*}\label{eq:lossAdv}
\mathcal{J} = \sum_{j=1}^M \| (\Psi_k)_j - (\widetilde{\Psi}_k)_j \|_2.
\end{equation*}
The parameters of $\ten$-MLP minimize the mean squared error between the true tension and the one reconstructed using the approximated Fourier coefficients. 

All the MLPs we use consist of \textit{fully connected layers} that are defined as follows. Let $W_k$ and $b_k$ be the $k^{th}$ fully connected layer's weight matrix of size $n_{k+1}$-by-$n_k$ and bias vector of length $n_k$. Then the layer transforms an input vector $z_{k}$ of length $n_k$ into a vector $W_k z_k + b_k$ of length $n_{k+1}$. In the MLPs we use \textit{batch normalization} after every fully connected layer except the last one to shift $W_k z_k + b_k$ to zero mean and scale it to unit variance over the data set. Finally, a nonlinear function $f$ computes an output $z_{k+1} = f(W_k z_k + b_k)$. Our choice for $f$ is the so-called \textit{leaky rectified linear unit} (leaky ReLu), i.e., $f(z) = \beta z$ if $z<0$, $f(z) = z$ if $z\geq 0$, where $\beta$ is a small constant. We set $\beta = 0.1$. The function $f$ operates on $W_k z_k + b_k$ elementwise. A MLP consists of a cascade of such transformations. A $K$-layer MLP maps an input $z_1$ to an output $z_{K+1}$ as follows
\begin{equation*}
z_2 = f(W_1 z_1 + b_1), \quad z_3 = f(W_2 z_2 + b_2), \quad \cdots, \quad z_{K+1} = f(W_K z_K + b_K).
\end{equation*}
The length of the vector $z_{k+1}$ is called the width of the $k^{th}$ layer. A MLP's parameters are the entries of the weight matrices $W_k$ and the bias vectors $b_k$. We use six fully connected layers in $R$-MLP. The input vector is of length 16 and the widths of the layers are 48, 96, 128, 256, 128 and 16. Therefore, the MLP has 85,920 parameters. $V$-MLPs and $T$-MLPs have the same architecture which consists of five fully connected layers. The widths of the layers are 48, 96, 128, 256 and 96. The MLP has 75,632 parameters. $\ten$-MLP has five fully connected layers. The widths of the layers are 64, 128, 256, 128 and 64 which make up 83,584 parameters. We normalize the inputs such that they have a unit mean and zero standard deviation.

Finding parameters minimizing the loss functions requires solving a nonconvex optimization problem. We solve this problem using a stochastic gradient descent method (SGDM). SGDM computes the gradients of a loss function with respect to the parameters of a MLP over a randomly sampled subset of the data set which is called a minibatch. In the trainings we use minibatches of size 256. SGDM iterations are terminated either when the error stagnates or SGDM exhausts the entire data set 20 times (i.e., after 20 epochs). SGDM has a parameter called learning rate that scales the size of a step taken in the descent direction. We set the initial learning rate to $5\times10^{-4}$ and scaled it by a factor of $0.2$ after ten epochs.

\begin{algorithm}[H]
\begin{algorithmic} 
\REQUIRE {$\xx_0, \tau, \theta, \mathtt{index}$}
\STATE {$\xx(\mathtt{index}) = \xx_0$}
\COMMENT {First re-order points}
\STATE {$\xx = \mathtt{rotate}(\xx,-\theta)$}
\COMMENT {Rotate back by $-\theta$}
\STATE {$\xx = \xx - \tau$}
\COMMENT {Finally, undo translation}
\end{algorithmic}
\caption{$\xx = \mathtt{destandardize}$, undoes ordering points, rotating and translating a vesicle} \label{a:destandardize}
\end{algorithm}

\begin{algorithm}[H]
\begin{algorithmic} 
\REQUIRE {$\uback$, time horizon $T$, spatial resolution $N_{LR}$ to take a step with the numerical scheme, initial vesicle shape $\xx^0$}
\STATE {$(N_{MLARM}, \Delta t, \kappa_b) \leftarrow (96, 10^{-4}, 1)$}
\COMMENT {Set MLPs' parameters}
\STATE {$(t, \xx) \leftarrow (0, \xx^0)$}
\COMMENT {Initialize time step counter and vesicle}
\WHILE {$t \leq T/\Delta t+1$}
\STATE {$ t = t + 1 $}
\COMMENT {Update time step counter}
\STATE {$[\mathtt{area},\mathtt{length}] = \mathtt{findAreaLength}(\xx)$}
\COMMENT {Find vesicle's area and arclength}
\IF[At every ten time steps switch to low-fidelity numerical scheme]{$\mathtt{rem}(t/10) == 0$}
\STATE {$\xx \leftarrow \mathtt{downSample}(\xx,N_{LR})$}
\COMMENT {Downsample $\xx$ to $N_{LR}$ points}
\STATE {$\xx^{\Delta t} = \mathtt{timeStep}({\xx},\kappa_b,\Delta t,\uback)$}
\COMMENT {Take time step by solving~\eqref{eq:aVesIE} using the numerical scheme with $N_{LR}$ points}
\STATE {$\xx^{\Delta t} \leftarrow \mathtt{upSample}(\xx^{\Delta t},N_{MLARM})$}
\COMMENT {Upsample to $N_{MLARM}$}
\ELSE[Otherwise take time step with MLPs]
\STATE {// First, solve advection problem (2)}
\STATE {$[\tau,\theta,\mathtt{index}] = \mathtt{findStandard}(\xx)$}
\COMMENT {Find standardization using Algorithm~\ref{a:findStandard}}
\STATE {$\xx = \mathtt{standardize}(\xx, \tau,\theta,\mathtt{index})$}
\COMMENT {Standardize $\xx$ using Algorithm~\ref{a:standardize}}
\STATE {$c = \WW^T_{16}\xx$}
\COMMENT {Find PCA coefficients $c$ of $\xx$}
\STATE {$\{\Psi_k\}_{k=1}^{N_f} = \mathtt{VMLP}(c)$}
\COMMENT {Approximate $V[\gamma]\phi_k$ for $k = 1, \cdots, N_f$ using $V$-MLPs}
\STATE {$\uback = \mathtt{standardize}(\uback(\xx),0,\theta,\mathtt{index})$}
\COMMENT {Evaluate $\uback$ at $\xx$ and standardize}
\STATE {$\{\huback\}_{k=1}^{N_f} = \mathtt{fft}(\uback)$}
\COMMENT {Find Fourier coefficients of $\uback$}
\STATE {$V[\gamma]\uback = \mathtt{upsample}(\sum_k^{N_f} \Psi_k \huback,N_{MLARM})$}
\COMMENT {Compute $V[\gamma]\uback$ and upsample to $N_{MLARM}$}
\STATE {$V[\gamma]\uback = \mathtt{destandardize}(V[\gamma]\uback,0,\theta,\mathtt{index})$}
\COMMENT {Destandardize $V[\gamma]\uback$ using Algorithm~\ref{a:destandardize}}
\STATE {$\xx = \xx + \Delta t V[\gamma]\uback$}
\COMMENT {Update $\xx$ due to $\uback$ as in (2)}
\STATE {// Second, solve bending problem (3)}
\STATE {$[\tau,\theta,\mathtt{index}] = \mathtt{findStandard}(\xx)$}
\COMMENT {Find standardization}
\STATE {$\xx = \mathtt{standardize}(\xx, \tau,\theta,\mathtt{index})$}
\COMMENT {Standardize $\xx$ }
\STATE {$c = \WW^T_{16}\xx_0$}
\COMMENT {Find PCA coefficients $c$ of $\xx$}
\STATE {$c^{\Delta t} = \mathtt{RMLP}(c)$}
\COMMENT {Approximate PCA coefficients of $\xx^{\Delta t}$ using $R$-MLP}
\STATE {$\xx^{\Delta t} = \WW_{16}c^{\Delta t}$}
\COMMENT {Construct $\xx^{\Delta t}$ from PCA coefficients}
\STATE {$\xx^{\Delta t} = \mathtt{destandardize}(\xx^{\Delta t},\tau,\theta,\mathtt{index})$}
\COMMENT {Destandardize $\xx^{\Delta t}$}
\ENDIF
\STATE{// Improve accuracy of $\xx^{\Delta t}$ using correction algorithms}
\STATE {$\yy^{\Delta t} = \mathtt{correctShape}(\xx^{\Delta t},\mathtt{area},\mathtt{length})$}
\COMMENT {Correct errors in area and arclength}
\STATE {$\xx^{\Delta t} = \mathtt{alignShape}(\yy^{\Delta t},\xx^{\Delta t})$}
\COMMENT {Align corrected shape's angle and center with $\xx^{\Delta t}$}
\STATE {$\yy^{\Delta t} = \mathtt{equallyDistributeInArcLength}(\xx^{\Delta t})$}
\COMMENT {Equally distribute points along arclength}
\STATE {$\xx^{\Delta t} = \mathtt{alignShape}(\yy^{\Delta t},\xx^{\Delta t})$}
\STATE {$\xx \leftarrow \xx^{\Delta t}$}
\COMMENT {Save new shape $\xx^{\Delta t}$ and move to next time step}
\ENDWHILE
\end{algorithmic}
\caption{$\mathtt{MLARM}$: Scheme for a single vesicle in a free-space flow} \label{a:NNAAlgo}
\end{algorithm}

\subsubsection{MLARM: Pseudo-algorithm}

In~\algref{a:NNAAlgo} we present the pseudo-algorithm of MLARM only for a vesicle in a free-space flow case due to its simplicity. As mentioned in the letter, we use several correction algorithms in order to improve accuracy of the solution and also stability of MLARM. One of them is the area and length correction algorithm ($\mathtt{correctShape}$). After correcting the errors in a vesicle's area and arclength, the vesicle might rotate and translate. We remove such artificial rotation and translation using $\mathtt{alignShape}$ algorithm. See~\cite{kabacaoglu-biros-e18} for the details of these algorithms. In the flows with multiple vesicles and solid walls, we compute the pairwise interactions between the boundaries using $N_{LR}$ points to discretize a vesicle membrane. So, the pairwise interactions are computed at a low resolution.


\subsection{Additional numerical results}

In this section, we explain the setup for the numerical experiments presented in the letter. Then we present results for a single vesicle in a free-space shear flow, the same-cost simulations of a single vesicle in a parabolic flow and discuss how the simulations capture the equilibrium lateral position of a vesicle in this flow. We also present the statistics of vesicles in dilute and dense Taylor-Couette flows and compare the MLARM simulations with the same-cost and the same-DOF ones.

\subsubsection{Setup for numerical experiments}\label{s:setup}

The true solutions are obtained using $N = 96$ points per vesicle and the time step size $\Delta t = 10^{-4}$. In the confined flow examples, we use 192 points to discretize the walls in the MLARM, the same-DOF, and the same-cost simulations and 256 points in the true solution. In the single vesicle examples (parabolic and shear flows, flow with curved flow lines), we use 16 PCA modes in MLARM and $N_{LR} = 16$ to solve the equations exactly at every ten time steps. In the confined flow examples, we use 32 PCA modes in order to have more degrees-of-freedom in vesicle shapes and $N_{LR} = 32$ to compute the pairwise interactions via FMM. In these examples, we do not solve the equations exactly at all. Accordingly, the number of points per vesicle is $N = 16$ in the same-DOF simulations of the single vesicle examples and $N = 32$ in those of the confined flow ones. The same-DOF simulations have $\Delta t = 10^{-4}$. We determine $N$ and $\Delta t$ for the same-cost simulations by taking several time steps with the numerical scheme for different $N$ and $\Delta t$ and matching the computation time. We set $N = 48$ and $\Delta t = 3\times10^{-4}$ in the same-cost simulations as they have the same CPU time as the MLARM simulations for this resolution.

\begin{figure}[!htb]
\centering
\begin{minipage}{0.5\textwidth}
\setcounter{subfigure}{0}
\centering
\renewcommand*{\thesubfigure}{(a)} 
\hspace{0cm}\subfigure[Comparison with the same-DOF simulations]{\scalebox{0.4}{{\includegraphics{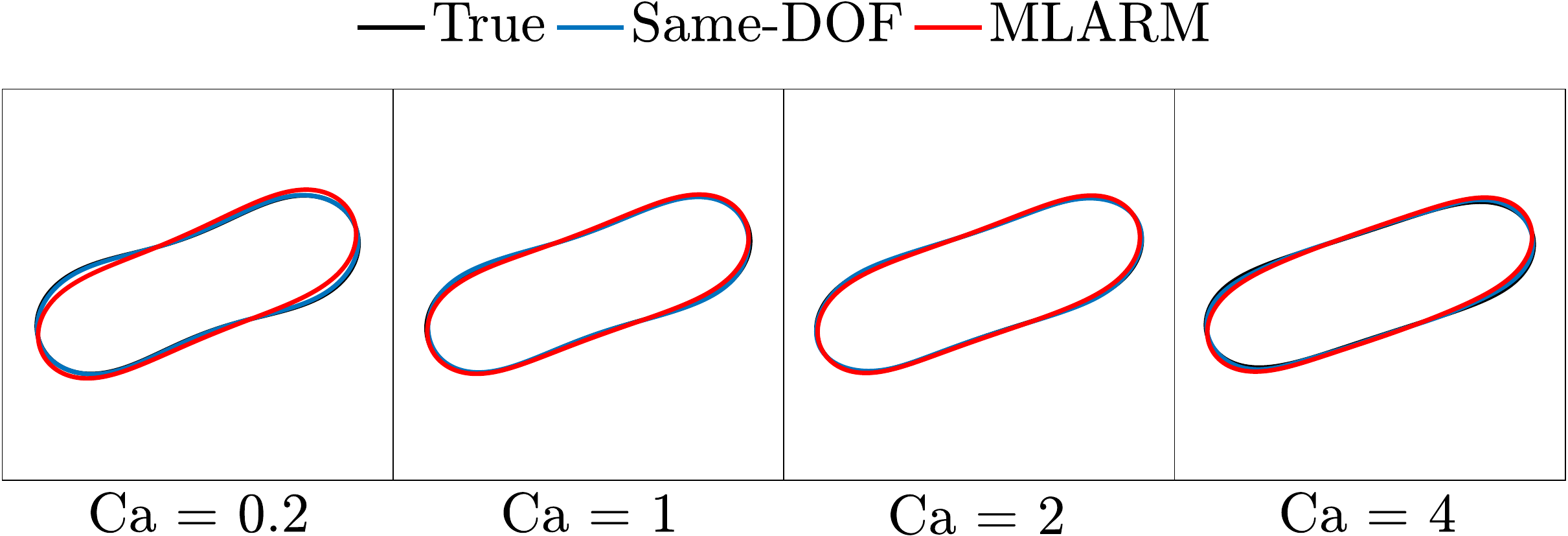}}}
\label{f:shearDOF}} 
\end{minipage}
\begin{minipage}{0.5\textwidth}
\setcounter{subfigure}{0}
\centering
\renewcommand*{\thesubfigure}{(b)} 
\hspace{0cm}\subfigure[Comparison with the same-cost simulations]{\scalebox{0.4}{{\includegraphics{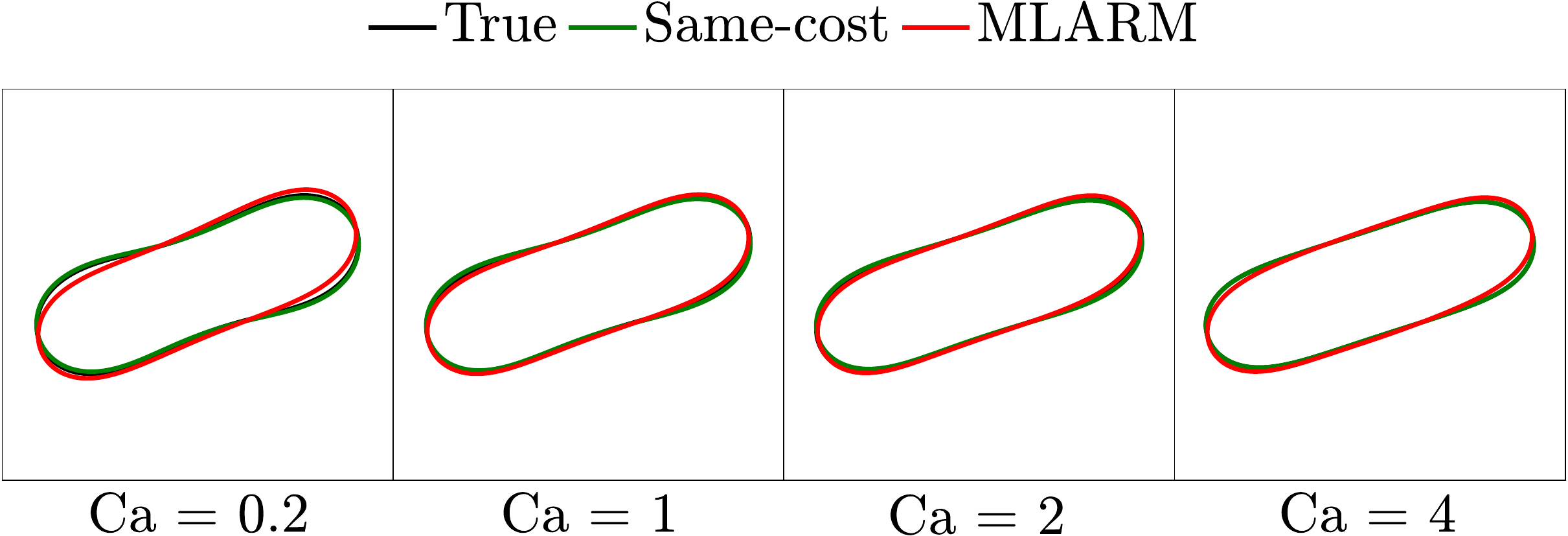}}}
\label{f:shearShapes}} 
\end{minipage}
\caption{MLARM vs. the same-DOF (a) and the same-cost (b) simulations for shear flows. Vesicles with no viscosity contrast tilts to a certain angle in such flows. The MLARM simulations accurately capture the equilibrium shapes and orientations. See~\figref{f:phaseAngle} for the tank-treading of the vesicle for Ca = 2.}\label{f:shearComps}
\end{figure}

\begin{figure}
\includegraphics[width=0.4\textwidth]{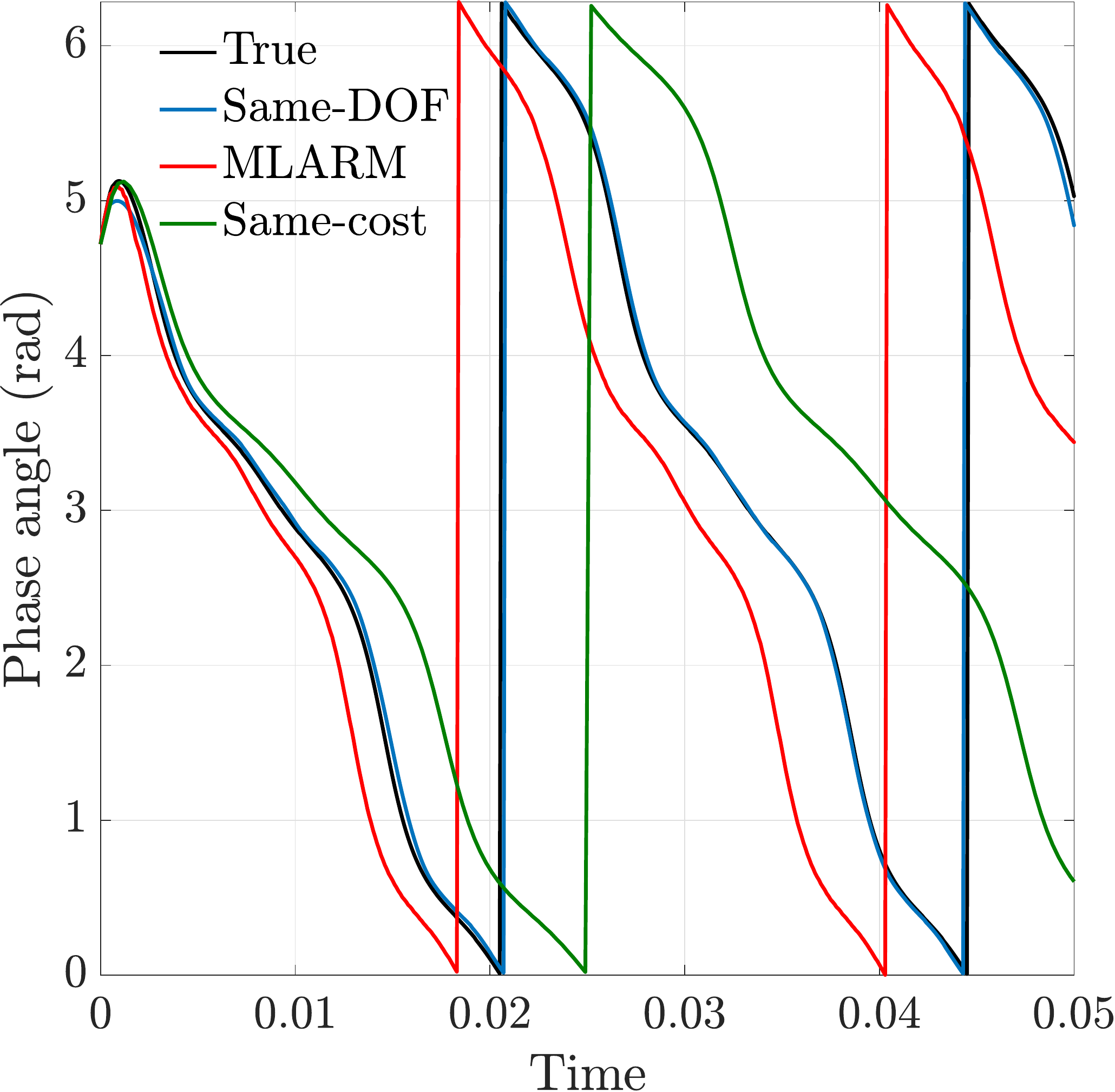}
\caption{Phase angle of the tank-treading vesicle in shear flow with Ca = 2. We plot the angle between the vesicle's principal axis and the vector from the vesicle's center to a specific point on the vesicle membrane.}
\label{f:phaseAngle}
\end{figure}

\subsubsection{Free-space shear flow}

We choose an initial vesicle shape and simulate it in shear flows ($\uback = \dot{\gamma}y\mathbf{e}_x$) for different Ca values. Since the fluids in the interior and the exterior of a vesicle have the same viscosity, the vesicle tilts to a certain angle and its membrane tank-treads~\cite{kraus-lipowsky-e96}. Figure~\ref{f:shearComps} shows the equilibrium shapes of the vesicle and compares the MLARM simulations with the same-DOF and the same-cost ones. Both low-resolution simulations and the MLARM simulations are similarly accurate in the equilibrium vesicle shapes. In order to investigate the vesicle's tank-treading we also present the vesicle's phase angle as a function of time in~\figref{f:phaseAngle}, i.e., the angle between its principal axis and the vector from the its center to a specific point on the membrane. The figure shows that although the MLARM simulation captures tank-treading, the vesicle tank-treads faster than the true solution does. Additionally, the same-cost simulation cannot accurately capture the tank-treading velocity either. The error stems from the numerical error introduced by the greater time step size in the same-cost simulation (which is $\Delta t = 3\times10^{-4}$) compared to the same-DOF and the MLARM simulations (which is $\Delta t = 10^{-4}$).

\begin{figure}[!htb]
\centering
\begin{minipage}{0.45\textwidth}
\setcounter{subfigure}{0}
\centering
\renewcommand*{\thesubfigure}{(a)} 
\vspace{0.4cm}\hspace{-0.5cm}\subfigure[Equilibrium shapes]{\scalebox{0.35}{{\includegraphics{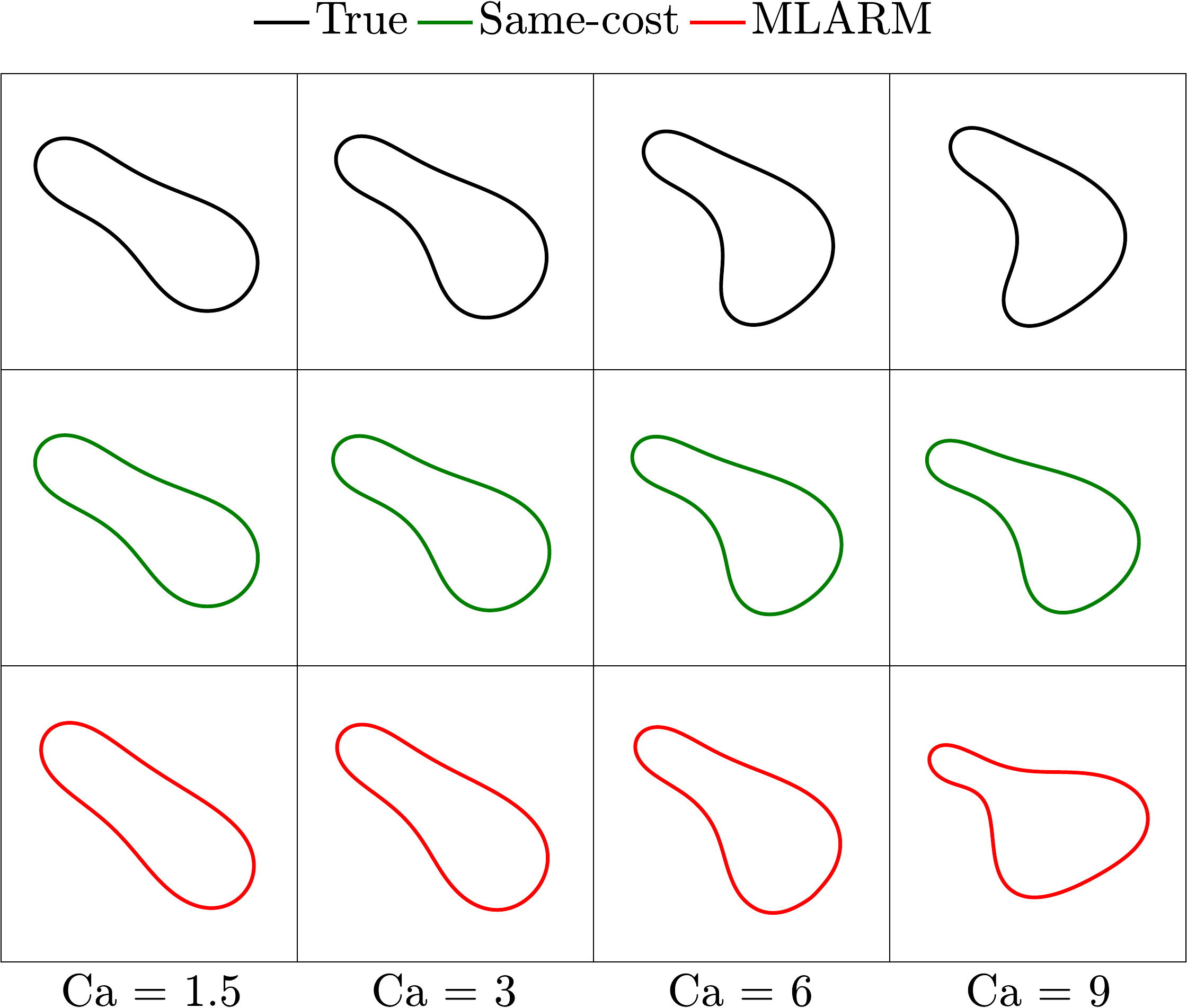}}}
\label{f:parabolShapes}} 
\end{minipage}
\begin{minipage}{0.45\textwidth}
\setcounter{subfigure}{0}
\centering
\renewcommand*{\thesubfigure}{(b)} 
\hspace{0.5cm}\subfigure[Equilibrium lateral positions]{\scalebox{0.35}{{\includegraphics{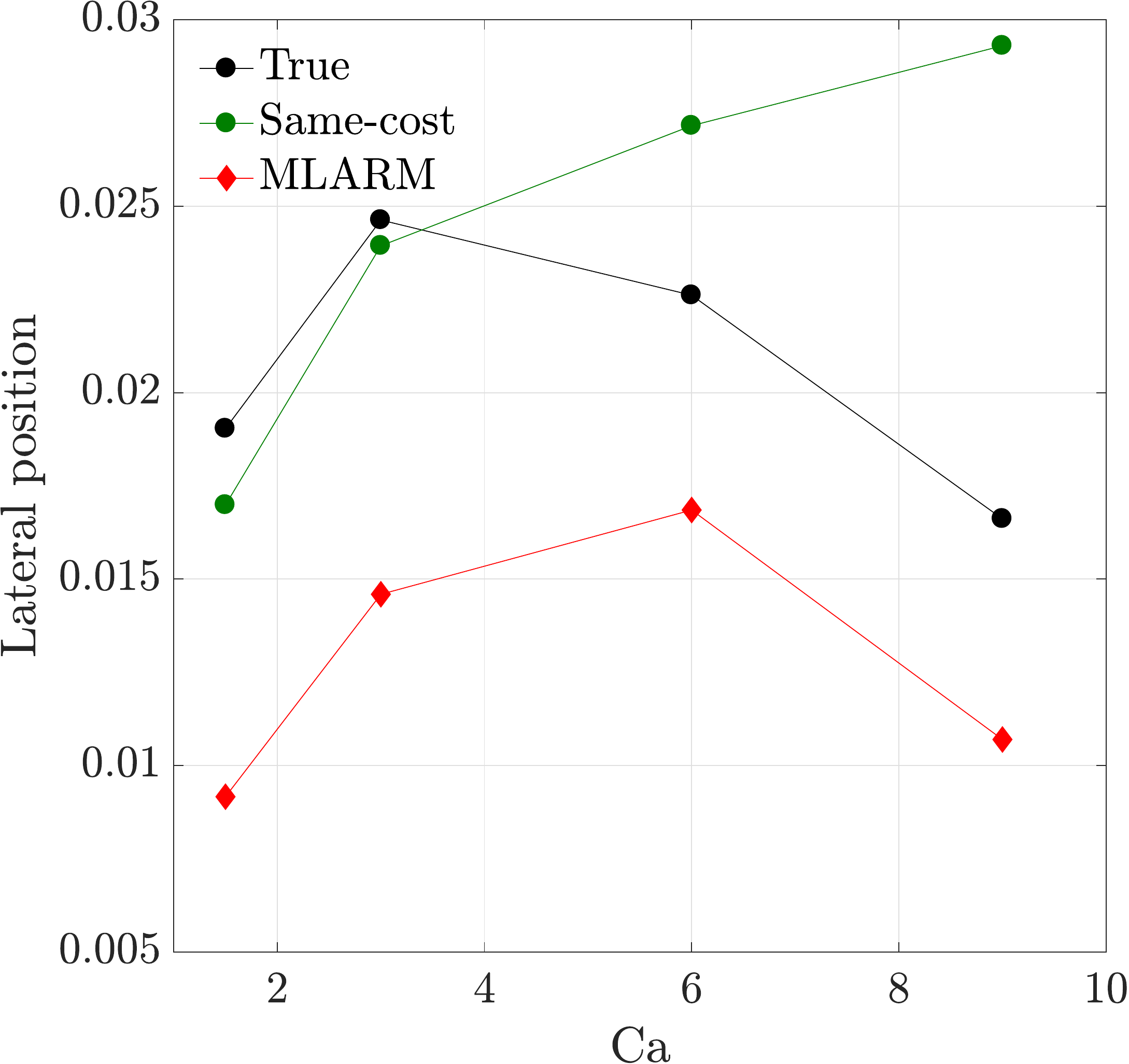}}}
\label{f:equilLatPos}} 
\end{minipage}
\caption{MLARM vs. the same-cost simulations for parabolic flows. (a) Equilibrium vesicle shapes in parabolic flows with different capillary numbers. (b) Equilibrium lateral positions of vesicles shown in (a). We omit the results for the same-DOF simulations since they deliver equilibrium lateral positions above 0.6 for all Ca values.}
\label{f:parabolicSameCost}
\end{figure}

\subsubsection{Free-space parabolic flow}

Figure 2 compares the same-DOF and the MLARM simulations in terms of capturing the equilibrium shapes of a vesicle in parabolic flows with different shear rates. Here we present the equilibrium shapes given by the same-cost simulations as well (Fig.~\ref{f:parabolicSameCost}). They are more accurate than the same-DOF ones in capturing the equilibrium shapes. However, they are not superior to the MLARM simulations. 

We also compare the equilibrium lateral positions of vesicles in these flows (Fig.~\ref{f:parabolicSameCost}). The MLARM simulations capture the equilibrium lateral positions more accurately than the same-cost ones, especially for high Ca.  We omit the same-DOF simulations as they have errors greater than 100\%.

\begin{figure}
\begin{minipage}{\textwidth}
\setcounter{subfigure}{0}
\renewcommand*{\thesubfigure}{(a)} 
\hspace{-0.5cm}\subfigure[AF = 2.2\%]{\scalebox{0.35}{{\includegraphics{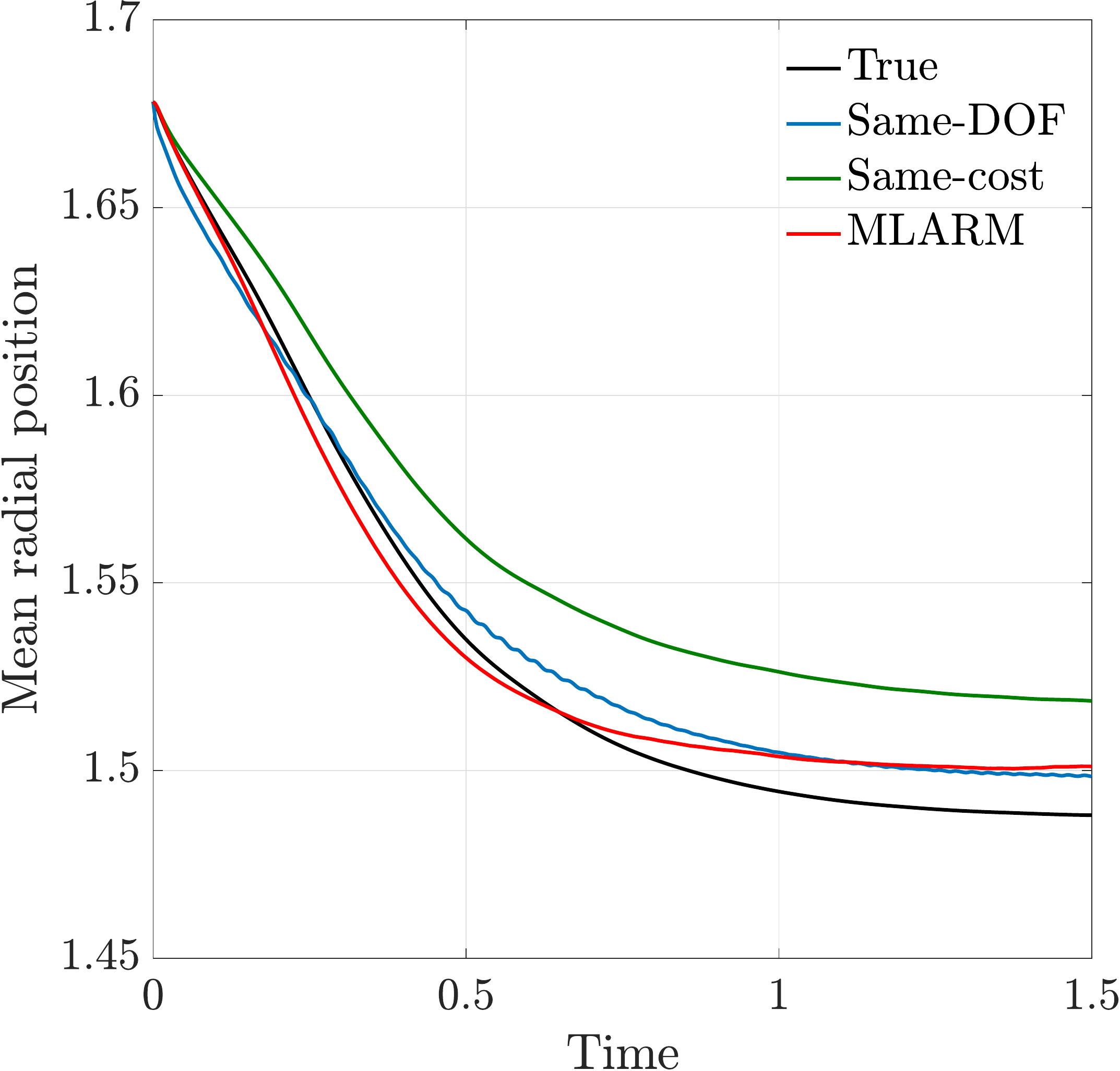}}}
\label{f:nv5meanRad}} 
\setcounter{subfigure}{0}
\renewcommand*{\thesubfigure}{(b)} 
\hspace{0.5cm}\subfigure[AF = 4.4\%]{\scalebox{0.35}{{\includegraphics{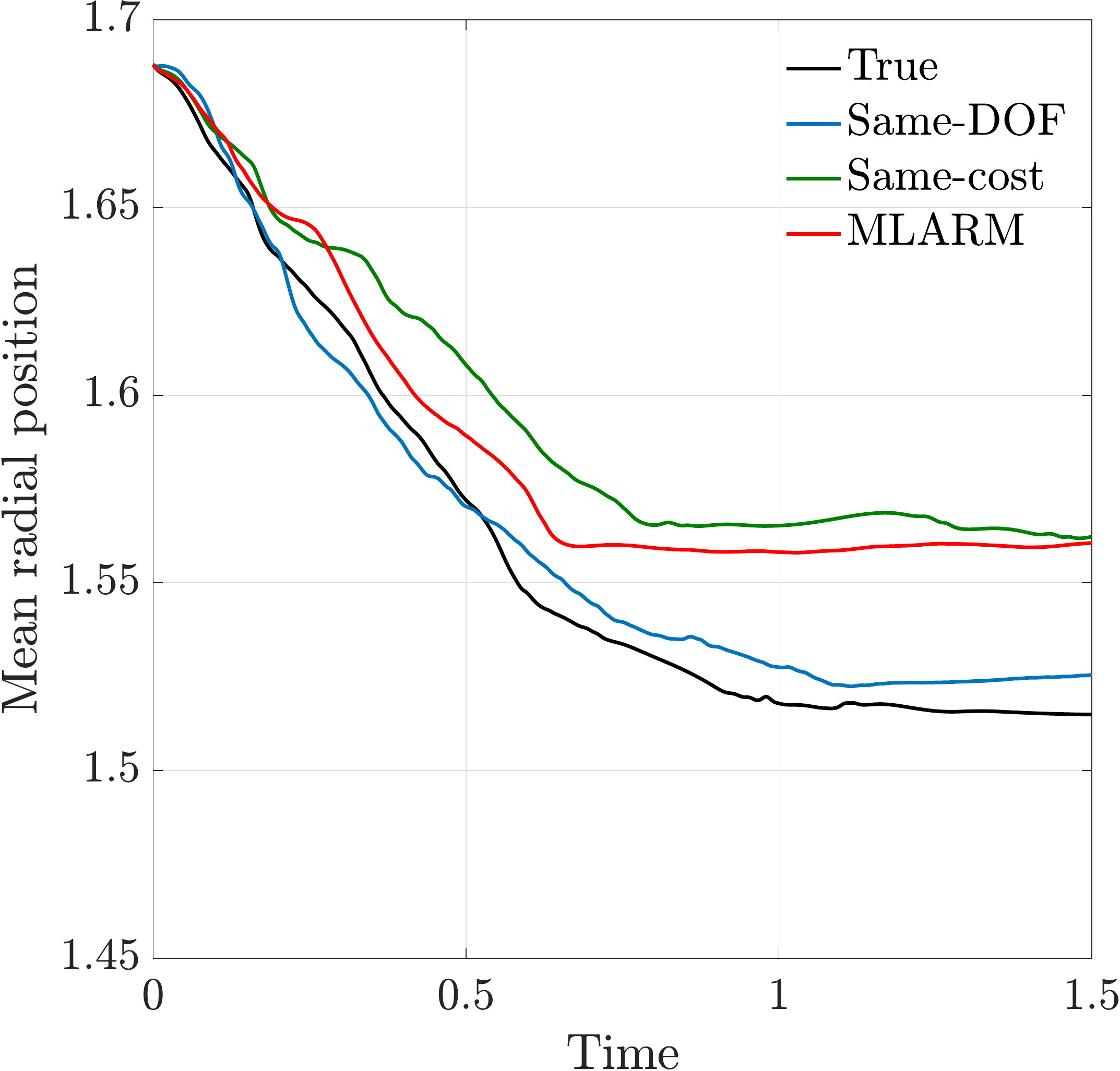}}}
\label{f:nv10meanRad}} 
\end{minipage}
\caption{Evolution of the mean radial position of vesicles in dilute Taylor-Couette flows.}
\label{f:tayCouette}
\end{figure}
\subsubsection{Dilute Taylor-Couette flows}
Figure 4 shows that vesicles in a dilute Taylor-Couette flow exhibit a spatial order by rotating in a rim with a uniform angular interdistance. We present the evolution of the mean radial position of the vesicles (i.e., the mean of the distances of vesicles' centers to the inner circle's center) for area fractions 2.2\% and 4.4\% in~\figref{f:tayCouette}. The figures show that the mean radial position converges as expected. The MLARM and the same-DOF simulations capture the true mean radial position accurately in the 2.2\% case. The MLARM simulation has a greater error in the 4.4\% case compared to the 2.2\% case. The reason is that MLARM ignores near interactions between vesicles and these interactions start dominating the dynamics as area fraction increases. For the denser case, the same-DOF simulation becomes more accurate than the MLARM simulation. 

In comparison with the same-cost simulation, the MLARM simulation is more accurate for an area fraction 2.2\%. The reason is that the same-cost simulation has a coarser temporal resolution than the MLARM simulation and capturing vesicle migration accurately requires fine temporal resolution. For an area fraction 4.4\%, however, the same-cost and the MLARM simulations have similar accuracy.

\begin{figure}[!htb]
\centering
\begin{minipage}{0.45\textwidth}
\setcounter{subfigure}{0}
\centering
\renewcommand*{\thesubfigure}{(a)} 
\hspace{-0.5cm}\subfigure[AF = 30\%]{\scalebox{0.28}{{\includegraphics{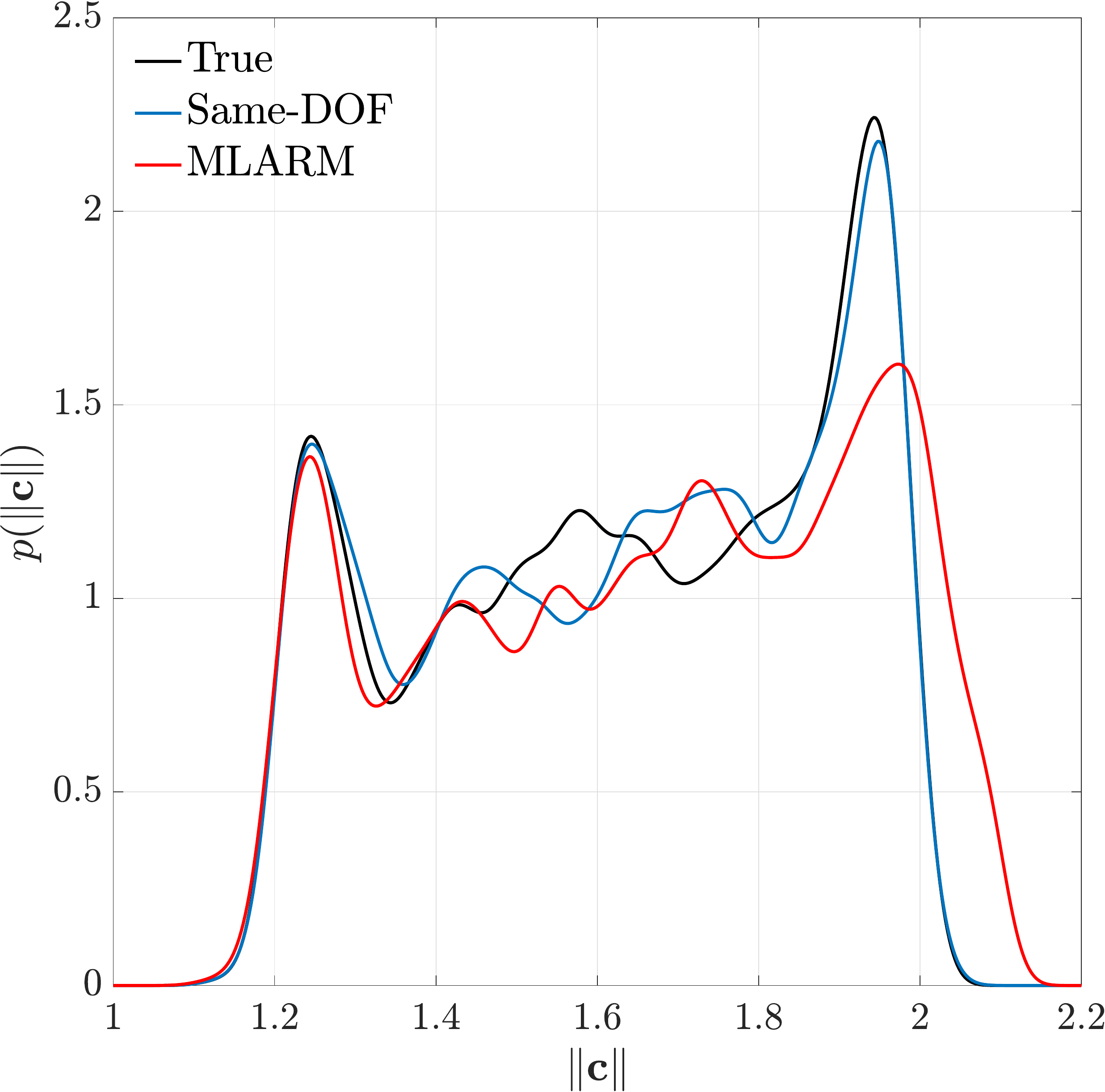}}}
\label{f:VF30stats}} 
\end{minipage}
\begin{minipage}{0.45\textwidth}
\setcounter{subfigure}{0}
\centering
\renewcommand*{\thesubfigure}{(b)} 
\hspace{0.5cm}\subfigure[AF = 35\%]{\scalebox{0.28}{{\includegraphics{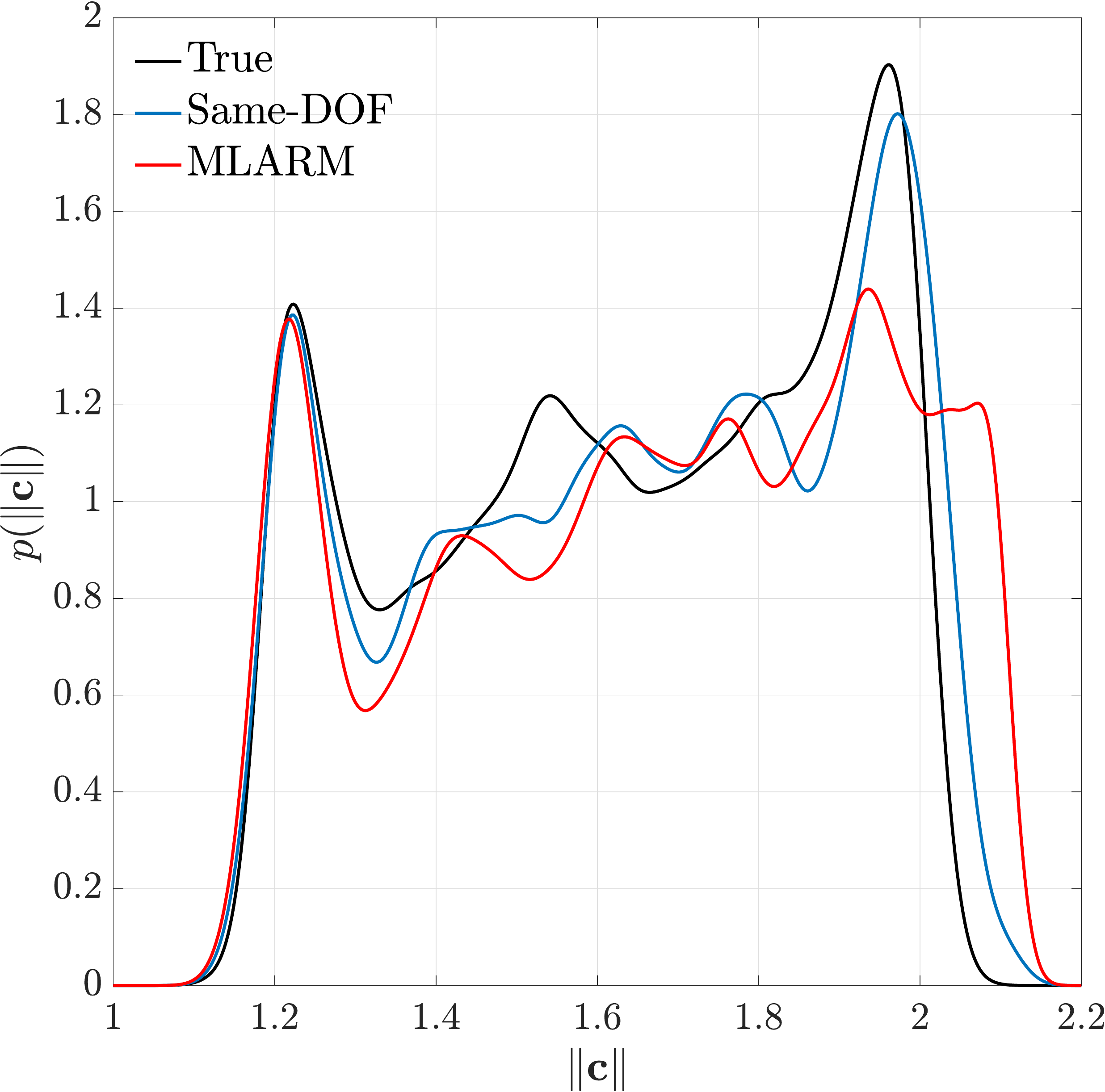}}}
\label{f:VF35stats}} 
\end{minipage}
\caption{Statistics of vesicles in dense Taylor-Couette flows. We plot the probability distribution of the distance of a vesicle's center to the origin $\| \mathbf{c}\|$ throughout the simulations. Vesicles form regions near circles that are free of vesicles. The MLARM simulations accurately capture these regions.}\label{f:denseStats}
\end{figure}
\subsubsection{Dense Taylor-Couette flows}
Figure 5 presents the probability distribution of the distance of a vesicles’ centers to the inner circle's center throughout the simulations at an area fraction 20\%. Here we present the results for area fractions 30\% and 35\% (\figref{f:denseStats}). The MLARM and the same-DOF simulations accurately capture the cell-free layers whereas there is no resolution that provides stable and the same computational simulations for 30\% and 35\% area fractions.

\end{widetext}

\end{document}